\documentclass[a4paper,reqno]{amsart}
\usepackage[text={5.0in,8in},centering]{geometry}
\usepackage{amsrefs}
\usepackage{amssymb}
\usepackage{mathrsfs}

\usepackage{xspace}

\usepackage{pgf}
\usepackage{tikz}

\usetikzlibrary{arrows,automata}

\usetikzlibrary{calc}

\DeclareMathAlphabet{\mathpzc}{OT1}{pzc}{m}{it}

\usepackage{epsfig}
\usepackage{enumerate}
\usepackage{graphicx}

\numberwithin{equation}{section}
\theoremstyle{plain}
\newtheorem{theorem}{Theorem}[section]

\newtheorem{lemma}[theorem]{Lemma}

\theoremstyle{remark}
\newtheorem{remark}{Remark}[section]

\newtheorem*{quest*}{Question}
\newtheorem*{remark*}{Remark}
\theoremstyle{remark}

\theoremstyle{definition}
\newtheorem{definition}{Definition}[section]
\newtheorem*{definition*}{Definition}
\newtheorem*{notation*}{Notation}
\newtheorem*{notations*}{Notations}

\providecommand{\D}{\mathbb}

\providecommand{\R}{\mathrm}

\newcommand{\ee}{\mathrm{e}}
\newcommand{\eul}{\mathrm{e}}

\def\ball{\mathrm{B}}
\def\bball{\mathbf{B}}

\DeclareMathOperator{\const}{const}
\DeclareMathOperator{\dist}{dist}

\DeclareMathOperator*{\essup}{ess\,sup}
\DeclareMathOperator*{\supp}{supp}

\DeclareMathOperator{\one}{\mathbf{1}}

\DeclareMathOperator{\mes}{{\rm mes}}

\def\tabhigh#1{{\buildrel _{} \over {#1}}}

\def\rc{\mathrm{c}}

\def\lam{{\lambda}}

\def\eps{\epsilon}

\def\CNR{{$E${\rm-CNR }}}

\def\LOC#1{\textsf{LOC}$(#1)$}

\def\SS#1{\textsf{S}$(#1)$}

\def\ER{$E$-R\xspace}

\def\CNR{$E$-CNR\xspace}
\def\PR{$E$-PR\xspace}

\def\EmS{$(E,m)$-S\xspace}
\def\EmNS{$(E,m)$-NS\xspace}

\def\ET{$E$-T\xspace}
\def\ENT{$E$-NT\xspace}
\def\ER{$E$-R\xspace}
\def\ENR{$E$-NR\xspace}

\def\tE{{ \widetilde{E} }}

\def\tf{{ \tilde{f} }}
\def\htil{{\widetilde{h}}}

\def\BbS{\mathbf{S}}

\def\DA{\D{A}}
\def\DC{\D{C}}

\def\DP{\D{P}}
\def\DR{\D{R}}
\def\DZ{\D{Z}}
\def\DN{\D{N}}

\def\cA{\mathcal{A}}
\def\cB{\mathcal{B}}
\def\csB{\mathscr{B}}

\def\cC{\mathcal{C}}
\def\cD{\mathcal{D}}
\def\cE{\mathcal{E}}
\def\cF{\mathcal{F}}
\def\cG{\mathcal{G}}

\def\cM{\mathcal{M}}
\def\cN{\mathcal{N}}
\def\cP{\mathcal{P}}
\def\csP{\mathscr{P}}
\def\csQ{\mathscr{Q}}

\def\cR{\mathcal{R}}
\def\cS{\mathcal{S}}

\def\cZ{\mathcal{Z}}

\def\cR{{\mathcal{R}}}
\def\cA{{\mathcal{A}}}
\def\cN{{\mathcal{N}}}

\def\boxx{{\mathrm B}}

\def\rD{{\R{D}}}
\def\rN{{\R{N}}}
\def\rd{{\R{d}}}

\def\rS{{\R{S}}}

\def\be{\begin{equation}}
\def\ee{\end{equation}}
\def\ba{\begin{array}{l}}
\def\ea{\end{array}}
\def\bal{\begin{aligned}}
\def\eal{\end{aligned}}

\def\fF{\mathfrak{F}}

\def\om{{\omega}}
\def\Om{{\Omega}}

\def\eps{\epsilon}

\def\Lam{{\Lambda}}
\def\lam{{\lambda}}

\def\pr#1{\D{P}\left\{\,#1\,\right\}}
\def\esm#1{\D{E}\left[\, #1\, \right]}
\def\pt{\partial}
\def\half{\frac{1}{2}}

\def\quart{\frac{1}{4}}

\def\nb#1{{ \langle #1 \rangle}}

\def\myset#1{{\left\{\,#1\,\right\}}}

\def\LOC#1{{\rm\textsf{LOC}}($#1$)\xspace}

\def\DeltaD{\Delta^{\rm D}}

\def\tabhigh#1{{\buildrel _{} \over {#1}}}

\begin{document}

\title[From fixed-energy MSA to dynamical localization]
{From fixed-energy MSA to dynamical localization: \\A continuing quest for elementary proofs}

\author[V. Chulaevsky]{Victor Chulaevsky}


\address{D\'{e}partement de Math\'{e}matiques\\
Universit\'{e} de Reims, Moulin de la Housse, B.P. 1039\\
51687 Reims Cedex 2, France\\
E-mail: victor.tchoulaevski@univ-reims.fr}

\date{}
\begin{abstract}
We review several techniques and ideas initiated by a remarkable work by Spencer
\cite{Sp88}, used and further developed in numerous subsequent researches. We also describe
a relatively short and elementary derivation of the spectral and strong dynamical
Anderson localization from the fixed-energy analysis of the Green functions,
obtained either by the Multi-Scale Analysis (MSA) or by the Fractional-Moment Method
(FMM). This derivation goes in the same direction as the Simon--Wolf criterion
\cite{SW86}, but provides quantitative estimates, applies also to multi-particle models and, combined with a simplified variant of the Germinet--Klein argument \cite{GK01}, results in an elementary proof of dynamical localization.
\end{abstract}

\maketitle

\section{Introduction. } \label{sec:intro}

The mathematical theory of Anderson localization, describing the motion a quantum particle (or
a collection of non-interacting particles) in a disordered environment has reached by now its
age of maturity. The number of different mathematical models (different even from the point of
view of applications to physical systems) and technical tools, allowing to analyze these models,
is quite impressive. On the other hand, this also makes  mastering these techniques
difficult for the beginners. It is not always easy to see the main ideas behind
dozens of pages filled with definitions, preliminary  facts and sophisticated arguments
pushed to the extreme due to the complexity of the problem. Yet, the maturity
of a mathematical theory can also be judged by the presence of comprehensive techniques
and simple general principles, guiding one through the jungle of more complex models and methods,
so a number of
leading researchers in this area of mathematical physics have been conducting a quest
for simpler, more elementary proofs of localization, intuitive (yet rigorous) techniques
and principles.

The quest began already in late 1980's, when Simon and Wolf
\cite{SW86} proved that suitable fixed-energy bounds on the Green functions
imply a.s. pure point spectrum, and then Spencer proposed in a remarkably short paper
\cite{Sp88} an elementary reformulation of the fixed-energy MSA developed in his pioneering
joint work with Fr\"{o}hlich \cite{FS83}. Besides the fact that  the paper
\cite{Sp88} came as a perfect complement to the Simon--Wolf
argument, it draws the reader's attention to the parallels between
the theory of random operators and a more
traditional spectral analysis of almost periodic operators, following the classical works
on periodic operators.

The first paper on the FMM, published by Aizenman and Molchanov \cite{AM93} in 1993, a few years
after the cycle \cite{FS83,FMSS85,SW86,Sp88,DK89}, also has been an important event in
the quest for elementary proofs (which the title of \cite{AM93} makes explicit). It took some time
and efforts to complete the first stage of the FMM (a fixed-energy analysis of fractional
moments of the resolvents) with additional arguments leading -- in a simple way -- to the strong dynamical localization. One can only regret that the MSA and the FMM have been evolving
in parallel with a very weak interaction, over the last two decades.

\vskip2mm

In the present paper, I am going to give a short and certainly incomplete
review of techniques and ideas brought to life or initiated indirectly by Tom Spencer
in his short article \cite{Sp88} which has the good fortune and privilege to
belong to those works which "\emph{tell much more than they say}".

The present text is intentionally left relatively short, to keep up with the spirit
of \cite{Sp88} (which was 10 page \emph{short}). For this reason, interesting applications
of the MSA to DSO with quasi-periodic (and more generally, deterministic) potentials had to be omitted;
these include deep analytic works
by Bourgain, Goldstein and Schlag (cf., e.g., \cite{BG00,BGS01}), where the MSA techniques
were applied to lattice Schr\"{o}dinger operators with "analytic" potentials,
and a recent article \cite{C11} where a different MSA-based approach has been used to treat
parametric families of almost periodic and some other \emph{deterministic} operators
by traditional -- and very simple -- methods of the theory of \emph{random} operators. On a personal note, I have to acknowledge with pleasure numerous fruitful discussions on this subject with Tom Spencer.

\vskip3mm

The pioneering work \cite{FMSS85} on Anderson localization in multidimensional disordered
media (following the work by Fr\"{o}hlich and Spencer \cite{FS83}) proved only the exponential  \emph{spectral }localization, i.e., pure point spectrum and exponential decay of generalized eigenfunctions. The relations between the spectral and dynamical manifestations of the Anderson localization phenomenon have been studied later in the work \cite{RJLS96} which has influenced
further development by Germinet--De Bi\`{e}vre \cite{GD98} and by Damanik--Stollmann \cite{DS01}.
The overall result of these researches (perfectly summarized by the title of the paper
\cite{DS01}) was a clear understanding that the variable-energy MSA (VEMSA, in short)
provides a sufficient input for the proof of the strong dynamical localization, in discrete
and continuous random media.

Germinet and Klein \cite{GK01} made a further step and gave a much shorter derivation of
the dynamical localization from the key MSA estimates, avoiding a tedious analysis of the
random geometry of the so-called centers of localization (the latter notion essentially goes
back to \cite{RJLS96}). The elegance of their approach resides, in particular, in the fact that the eigenfunction correlator bounds are inferred from those provided by the MSA directly in the entire configuration space (a Euclidean space, in their case). However, this elegance comes with a price: one has to rely upon a deep analysis of weighted Hilbert--Schmidt norms of spectral projections of Schr\"{o}dinger (or some other) operators in $L^2(\DR^d)$. Such an analysis has been carried out by Simon
\cite{Sim82} for potentials bounded from below by  $-C(|x|+1)$, and later extended by
Poerschke and Stolz \cite{TS93} to potentials bounded from below by $-C(|x|^2+1)$; the latter is a usual condition for the essential self-adjointness of the respective Schr\"{o}dinger operator.
In arbitrary \emph{finite} volumes, the analog of Germinet--Klein argument is reduced to a three-word instruction: "\emph{Apply Bessel's inequality}".

The situation is particularly simple for operators on a countable graph, where functional analysis is in fact replaced by linear algebra. Indeed, the required eigenfunction expansions are completely elementary here, for Hermitian operators in finite-dimensional Hilbert spaces. In continuous configuration spaces (Euclidean spaces and quantum graphs), a similar effect is achieved whenever the random operators in question have (as they usually do) compact resolvent in any finite volume, under reasonable requirements on its geometry. Physically speaking, only such finite-volume  bounds (uniform in the size of the volume) are relevant for applications to the quantum transport in disordered media. The term "finite volume" should not be misleading: a sample of a random media of the size of the Milky Way is still finite ... and the task of designing computer processors (let alone nano-devices) of \emph{that} size does not occupy yet the minds in the physics community.

Mathematically speaking, once uniform bounds on eigenfunction correlators are obtained in finite volumes of arbitrarily large size, one is just three words away from the strong dynamical localization in the entire space: "\emph{Apply Fatou lemma}" (cf., e.g., \cite{A94}, \cite{ASFH01}).

With these observations in mind, I propose here a streamlined derivation of the VEMSA-type probabilistic bounds from their simpler FEMSA counterparts.
Several elements of such a derivation appeared earlier, e.g., in \cite{MS85}, \cite{ETV10}
(but the approach from Section \ref{sec:SW.PCT} appears to be original).
It plays a role similar to that of the Simon--Wolf argument, providing a `soft way'
from the fixed-energy localization to stronger manifestations of the Anderson
localization phenomenon.
The key notion here is what is called in Sections \ref{sec:SW.ETV}--\ref{sec:SW.PCT}
the "singular width" of the spectrum:
the total Lebesgue measure of a (reasonably large) finite number of intervals
chosen in such a way that outside this "singular zone" Green functions
are exponentially small. (Curiously, the abbreviation "SW" suits to "Simon--Wolf",
"singular width" and "soft way"; the central symetry transforms it into
"MS", as in "Martinelli--Scoppola", cf. \cite{MS85}.)

\vskip2mm

The structure of this paper is as follows.
\begin{itemize}
  \item Main notions and notations are introduced in Section \ref{sec:notations}.
  \item In Sections \ref{sec:GRI}--\ref{sec:SH}, we give a streamlined version of the
  analytic core of the fixed-energy MSA, in the simplest form going back to
  \cite{Sp88}, but formulated in a more abstract fashion.
  \item In Section \ref{sec:fixed.energy}, we show that the simple, fixed-energy analysis
  from \cite{Sp88} can be easily improved so as to provide the key probabilistic bounds on the Green functions stronger than any power law.
  \item In Section \ref{sec:SW.ETV}, following essentially\footnote{I thank Tom Spencer and Sasha Sodin for a fruitful discussion of the works \cite{ETV10}--\cite{ESS12}.}
   \cite{ETV10}, we derive from fixed-energy bounds their variable-energy counterparts. The obtained results apply also to the FMM-type bounds (which are always initially obtained at fixed energy).

   \item In Section \ref{sec:SW.PCT}, we give another derivation of the variable-energy bounds which allows to prove directly the exponential spectral localization (i.e., the exponential decay of eigenfunctions), as well as the dynamical localization, under a stronger assumption on the random potential (cf. Eqn \eqref{eq:cond.PCT}).
  \item In Section \ref{sec:MSA.to.DL}, we formulate a finite-volume variant of the Germinet--Klein argument,
  allowing to derive the strong dynamical localization from variable-energy bounds
  on the resolvents (obtained by the MSA or by the FMM).
  \item Section \ref{sec:non.bootstrap} describes a simple adaptation of the techniques from Section \ref{sec:fixed.energy} sufficient for the proof of sub-exponential dynamical
      localization.
  \item In Section \ref{sec:DSA}, we briefly describe another simple approach
  (developed in our recent paper \cite{C12}) which also has its roots in \cite{Sp88}. However, the main object of the scale induction is here the decay of the eigenfunctions in finite
  balls, rather than the decay of Green functions.
\end{itemize}

The principal statements are theorems  \ref{thm:sing.width},
\ref{thm:SW.2vol}, \ref{thm:GK}, \ref{thm:subexp.ind.loc} and \ref{thm:fixed.subexp}.

For the sake of brevity and clarity of presentation, I do not discuss several
powerful (but more complex)
techniques from the works by Germinet--Klein, including the bootstrap MSA
(cf. \cite{GK01}) and spectral reductions from \cite{GK11} used in a very general framework of
random operators in $\DR^d$
with singular probability distributions.

\vskip2mm

To conclude the introduction, I would like to emphasize the role that the paper
\cite{Sp88} has played in a recent development of the multi-particle MSA (MPMSA).
In our joint works with Yuri Suhov (cf., e.g., \cite{CS09b}),
we aimed initially to prove the spectral
localization, which requires traditionally a variable-energy analysis. However,
the fixed-energy analysis has a substantial advantage to simplify both
geometrical and analytical ingredients of the MPMSA. I plan to address this subject
in a forthcoming work, using the reductions described in Sections \ref{sec:SW.ETV}--\ref{sec:MSA.to.DL}.

\section{Basic notations, facts and assumptions}
\label{sec:notations}

Throughout this paper, we work with discrete Schr\"{o}dinger operators (DSO) acting in
Hilbert spaces of square-summable complex functions on connected countable graphs.
Indeed, the techniques and results of the MSA, initially developed for operators on periodic lattices, are naturally extended to more general graphs with polynomially bounded growth of
balls (such graphs as Bethe lattices remain so far out of the MSA's reach). Another motivation for
presenting the new approach on a graph comes from the fact that the natural language
for the description of a system of $N>1$ interacting indistinguishable quantum particles
(bosons or fermions) is that of a symmetric power of the configuration space  $\cZ$ of the
respective single-particle system; already in the case where the configuration space
is $\cZ=\DZ^d$, $d>1$, its $N$-th \emph{symmetric} power is no longer a periodic lattice.

Consider a finite or countable connected graph $(\cG, \cE)$, with the set of vertices $\cG$ and the set of edges $\cE$; for brevity, we will often call $\cG$ the graph, omitting the reference to $\cE$. We denote by $\rd_\cG(\cdot,\,\cdot)$ (sometimes simply by $\rd(\cdot\,,\cdot)$)
the canonical distance on the graph $\cG$: $\rd_\cG(x,y)$ is the length of the shortest path
$x\rightsquigarrow y$ over the edges.
We will assume that the growth of balls $\ball_L(x):=\{y:\, \rd_\cG(x,y)\le L\}$ is polynomially bounded:
\be\label{eq:ball.growth}
\sup_{x\in\cG}| \ball_L(x)| \le C_d L^d, \;\; L\ge 1.
\ee
In particular, the coordination number $n_\cG(x) :=\{y:\, \rd_\cG(x,y)=1\}$ of any vertex $x$
is bounded by $C_d$ (even by $C_d - 1$).
The canonical (negative) graph Laplacian $(-\Delta_\cG)$ on a finite or countable graph
$(\cG, \cE)$ is given by
\be\label{eq:def.Laplace.graph}
(-\Delta_\cG f)(x) = \sum_{ \nb{x,y} } (f(x) - f(y))
= n_\cG(x) f(x) - \sum_{ \nb{x,y} } f(y)
\ee
where we use a popular notation $\nb{x,y}$ for a pair of nearest neighbors $x,y\in\cG$, i.e., $\rd_\cG(x,y)=1$, and $n_\cG(x)$
is the coordination number of the point $x$. For brevity, we will sometimes use
slightly abusive notations like $\nb{x,y}\in\Lam$, $\Lam\subset\cG$
instead of $\nb{x,y}\in(\Lam\times\Lam) \cap \cE_\cG$.

From this point on, unless otherwise specified, we will use the notation $\cG$ only for finite connected graphs, while $\cZ$ will stand for a countable connected graph
with polynomial growth of balls. In operator form, we can write
$$
-\Delta_\cG = n_\cG - \sum_{ \nb{x,y} } \Gamma_{x,y},
\quad
\Gamma_{x,y} = | \one_x \rangle \langle \one_y |,
$$
where $n_\cG$ is the operator of multiplication by the function $x\mapsto n_\cG(x)$.
Given a proper (connected) subgraph $\Lam\subsetneq \cG$, define its internal, external and
the so-called edge boundary (relative to $\cG$) as follows:
$$
\bal
\pt^-_{\cG} \Lam &= \{y\in\Lam:\, \rd_\cG(x, \cG\setminus \Lam) = 1\},
\quad
\pt^+_{\cG} \Lam = \pt^-_{\cG } \cG\setminus \Lam,
\\
\pt_{\cG} \Lam &= \{(x,y)\in\pt^-_\cG \Lam\times\pt^+_\cG  \Lam: \,\rd_\cG(x, y)=1\}.
\eal
$$

Working with a given graph $\cG (\subset\cZ)$, we always mean by a ball $\ball_R(u)\subset\cG$
the set $\{y\in\cG:\, \rd_\cG(u,y)\le R\}$, i.e., the \textbf{ball relative to the metric space}
$(\cG,\rd_\cG)$.

The Laplacian (hence, a DSO) in a subgraph $\Lam\subset\cG$ can be defined in various ways. The two most popular choices are:
\begin{itemize}
  \item The canonical (negative) Laplacian in $\Lam$, $(-\Delta^\rN_\Lam f)=(-\Delta_\Lam f)$,
  defined as in \eqref{eq:def.Laplace.graph} with $\cG$ replaced by
  $\Lam$. It this context, it is usually considered as an analog of the Neumann Laplacian, and reads as follows:
\be\label{eq:Laplace.N}
(-\Delta^\rN_\Lam f)(x) = n_{\Lam}(x) - \sum_{\nb{x,y}\in \Lam} f(y).
\ee
  \item The Dirichlet Laplacian
$(-\Delta^\rD_{\Lam,\cG}) = \one_{\Lam} (-\Delta^\rD_{\Lam,\cG}) \one_{\Lam} \upharpoonright
\ell^2(\Lam)$. Here we use a natural injection $\ell^2(\Lam)\hookrightarrow \ell^2(\cG)$.
The Dirichlet counterpart of \eqref{eq:Laplace.N} is
\be\label{eq:Laplace.N}
(-\Delta^\rD_\Lam f)(x) = n_{\cG}(x) - \sum_{\nb{x,y}\in \Lam} f(y),
\ee
with $n_\cG(x)\ge n_\Lam(x)$, so
$(-\Delta^\rD_\Lam) \ge (-\Delta^\rN_\Lam)$ in the sense of quadratic forms.
\end{itemize}

We will use the Dirichlet Laplacians and DSO $H^\rD_\Lam$.
Given a decomposition $\cG = \Lam \sqcup \Lam^\rc$, $\Lam^\rc := \cG\setminus\Lam^\rc$,
we can write
$$
\bal
-\DeltaD_\cG &= n_\cG -\sum_{ \nb{x,y}\in\Lam } \Gamma_{x,y} - \sum_{ \nb{x,y}\in\Lam^\rc } \Gamma_{x,y}
- \sum_{ \nb{x,y}\in\pt \Lam } \left(\Gamma_{x,y} + \Gamma_{y,x} \right)
\\
& =  \left( (-\DeltaD_\Lam) \oplus (-\DeltaD_{\Lam^\rc}) \right) - \Gamma_{\Lam,\cG}
\eal
$$
with $\Gamma_{\Lam,\cG}=\sum_{ \nb{x,y}\in\pt \Lam } \left(\Gamma_{x,y} + \Gamma_{y,x} \right)$.
Respectively for the DSO $H_\cG = -\DeltaD_\cG + V$, where
$V:\cG \to \DR$ is usually referred to as the potential, one has
$$
H_\cG
= H^\bullet_{\cG,\Lam} - \Gamma_{\Lam,\cG},
\qquad
H^\bullet_{\cG,\Lam} := (-\DeltaD_\Lam + V) \oplus (-\DeltaD_{\Lam^\rc} + V).
$$
We omit the superscript "$\rN$", since the nature of the boundary conditions in $\cG$
is not related to the choice of Dirichlet or Neumann decoupling
induced by $\cG = \Lam \sqcup \Lam^\rc$.

The spectrum of a (finite-dimensional) operator $H_\cG$, i.e., the set of its eigenvalues (EVs)
counted with multiplicities, will be denoted by $\Sigma(H_\cG)$.

In a number of formulae and statements, we will use the parameters $\beta,\tau,\varrho\in(0,1)$,
and $\alpha\in(1,2)$. Unless otherwise specified, we assume that $\beta=1/2$, $\tau=1/8$,
$\varrho = (\alpha-1)/2 = 1/6$ and $\alpha = 3/2$. Note that the exponent
$\frac{1+\varrho}{\alpha}$ figuring in Definition \ref{def:NR.NS} then equals $7/8$.
(These settings \textbf{will be changed in Section} \ref{sec:non.bootstrap}.)

\begin{definition}\label{def:NR.NS}
Given numbers $E\in\DR$ and $m>0$, a ball $\ball_L(u)$ is called
\begin{itemize}
  \item $E$-resonant (\ER, in short), if $\dist(\Sigma(H_{\ball_L(u)}), E) < \eul^{-L^\beta}$,
  and $E$-nonresonant (\ENR), otherwise;
  \item $(E,m)$-nonsingular (\EmNS), if for all $x,y\in\ball_L(u)$ with
  $\rd(x,y)\ge L^{\frac{1+\varrho}{\alpha}}$
\be\label{eq:def.NS}
C^2_d L^d \cdot
|G_{\ball_L(u)}(x,y;E)| \le \eul^{-\gamma(m,L)\rd(x,y)},
\ee
where
\be
\gamma(m,L) := m(1+L^{-\tau}),
\ee
and $(E,m)$-nonsingular (\EmNS), otherwise.
\end{itemize}

\end{definition}

Observe that for any ball $\ball_L(u)$, $|\pt \ball_L(u)| \le C^2_d L^d$, by virtue of
\eqref{eq:ball.growth}.

\subsection{Assumptions on the random potential}

For clarity of presentation, we always assume that the random potential field
$V:\cZ\times\Om\to\DR$ on a graph $\cZ$ is IID, with Lipshitz continuous
marginal probability distribution function (PDF) $F_V$:
\be\label{eq:Lipshitz}
\sup_{t\in\DR} (F_V(t+s) - F_V(t)) \le C_W s, \quad C_W\in(0,+\infty).
\ee
It is well-known that this assumption can often be relaxed to uniform H\"{o}lder continuity,
and even to a form of log-H\"{o}lder continuity.

\subsection{The Wegner estimate}

The original result by Wegner \cite{W81} has been adapted to a large number of
classes of random operators. Here we apply its simplest version, for DSO with a Lipshitz continuous IID random potential. The proof can be found in a number
of books and review articles; cf.,  e.g., Lemma VIII.1.8 in \cite{CL90}.

\begin{lemma}[Wegner estimate]\label{lem:Wegner}
Under the assumption \eqref{eq:Lipshitz}, for any finite graph $\cG$
of cardinality $|\cG|$ and any $\eps\in[0,1]$
\be\label{eq:Wegner}
\sup_{E\in\DR} \pr{ \dist(\Sigma(H_{\cG}), E) \le \eps} \le C_W |\cG| \eps.
\ee
\end{lemma}

In fact, the above statement remains valid for any ensemble of random operators
of the form $V(\,\cdot\,;\om) + H_0$ with a non-random operator $H_0$, for only the diagonal
part $V:\cG\times\Om\to\DR$ is used in the proof (cf. \cite{CL90}).

\section{Decoupling of resolvents on graphs}
\label{sec:GRI}

\subsection{Geometric resolvent inequality}

The second resolvent identity implies the so-called Geometric resolvent equation for the resolvents $G_\cG(E) = (H_\cG - E)^{-1}$, $G_{\Lam^\rc}(E) = (H_{\Lam^\rc} - E)^{-1}$,
$G^\bullet_{\cG,\Lam}(E) = (H^\bullet_{\cG,\Lam} - E)^{-1}$:
\be\label{eq:GRE.resolvents}
G_\cG(E) = G^\bullet_{\cG}(E) + G^\bullet_{\cG}(E) \, \Gamma_{\Lam,\cG} \, G_\cG(E).
\ee
For $x, u\in\Lam$ and $y\in\Lam^\rc$, one has
$G^\bullet_{\cG}(x,u;E)=G^\rD_{\Lam}(x,u;E)$ and
$G^\bullet_{\cG}(x,y;E)=0$. This results in the Geometric resolvent equation
for the Green functions
\be\label{eq:GRE}
G_\cG(x,y;E) = \sum_{ \nb{u,u'}\in\pt \cG \Lam} G^\rD_{\Lam}(x,u;E)\, G_{\cG}(u',y;E)
\ee
and the Geometric resolvent inequality (GRI)
\be\label{eq:GRI}
|G_\cG(x,y;E)| \le
\sum_{ \nb{u,u'}\in\pt \cG \Lam} |G^\rD_{\Lam}(x,u;E)|\, |G_{\cG}(u',y;E)|.
\ee

\section{Subharmonicity on graphs}
\label{sec:SH}

\subsection{Regular subharmonic functions}

\begin{definition}\label{def:subh}
Let $\cG$ be a finite connected graph, $L\ge \ell\ge 0$ two integers and $q\in(0,1)$.
A function $f:\cG\to\DR_+$ is called $(\ell,q)$-subharmonic in a ball $\ball_L(u)\subsetneq \cG$
if for any ball $\ball_\ell(x)\subseteq\ball_L(u)$ one has
\be\label{eq:lqsubh}
f(x) \le q \max_{y\in\ball_{\ell + 1}(x)} f(y).
\ee
\end{definition}

We will often use the notation $\cM(f,\Lam) := \max_{x\in\Lam}| f(x)|$.
\begin{lemma}\label{lem:subh.1}
If a function $f:\cG\to\DR_+$ defined on a finite connected graph $\cG$ is $(\ell,q)$-subharmonic
in a ball $\ball_L(x)\subsetneq\cG$, with $L\ge \ell\ge 0$, then
\be\label{eq:lem.subh.1}
f(x) \le q^{ \left\lfloor \frac{L+1}{\ell+1} \right\rfloor} \cM(f,\cG)
     \le q^{ \frac{L - \ell }{\ell+1} } \cM(f,\cG).
\ee
\end{lemma}

In fact, the factor $\cM(f,\cG)$ in the RHS of \eqref{eq:lem.subh.1} can be replaced
by $\cM(f,\ball_{L+1}(x))$.

\proof
Since $L\ge \ell\ge 0$, we have $n+1 := \left\lfloor \frac{L+1}{\ell+1} \right\rfloor\ge 1$.
Set $\Lam_j := \ball_{j(\ell+1)}(x)$, $0 \le j \le n$, and note that
$\Lam_{n+1} \subset\ball_{L+1}(x)$, since $(n+1)(\ell+1) \le L+1$. Further, if $y\in\Lam_j$
with $0\le j \le n$
and $z\in\ball_{\ell+1}(y)$, then, by triangle inequality, $z\in\Lam_{j+1}$. Consider a monotone non-decreasing function $h:[0,L+1]\cap \DN \mapsto \DR_+$ defined by
$h(r) = \cM(f, \ball_r(x))$. Using the $(\ell,q)$-subharmonicity of $f$, we obtain
$$
h( j(\ell+1) ) \le q \max_{y\in \Lam_j} \max_{z\in \ball_{\ell+1}(y)} f(z)
\le q h( (j+1)(\ell+1)),
$$
in particular,
$$
h( n(\ell+1)) \le q h(L+1) \le q \cM(f,\cG).
$$
Since $f(x) = h(0)$, the claim follows by the backward recursion in $j$ from $n$ to $0$,
in $n$ steps.
\qedhere
\vskip4mm

\textbf{Example.} $\cG = [0,R]\cap \DZ$, $R=L+1$, $\ell=0$, and $f:x\mapsto q^{R-x}$,
$x\in\cG$. For all $y\in \ball_L(0) = [0,L]$, one has
$$
f(x) = q^{R-x} = q \, \max_{|y-x|\le 1} q^{R-y}
=  q \, \max_{|y-x|\le 1} f(y),
$$
which implies the $(0,q)$-subharmonicity of the function $f$ in $\ball_L(0)$. In fact, here the inequality of the form \eqref{eq:lqsubh} turns out to be an equality, and one has
$$
f(0) = q^{L+1} = q^{\frac{L+1}{0+1}} f(L+1),
$$
which shows that the estimate from Lemma \ref{lem:subh.1} is sharp. Note also that
the inequality \eqref{eq:lqsubh} cannot be extended to the exterior point $y=L+1$, since
$$
f(L+1) = 1 > q = f(L).
$$
Clearly, a function $(\ell,q)$-subharmonic everywhere in $\cG$, with $q<1$,
must be zero:
$$
0 \le \max_x f(x) \le q \max_y f(y).
$$

Naturally, the main \emph{raison d'\^{e}tre} of the Definition \ref{def:subh}
is the following fact.

\begin{lemma}\label{lem:GF.is.SH.1}
Consider a finite connected graph $\cG$ and a ball $\ball_L(u)\subsetneq \cG$, with
$L\ge \ell\ge 0$. Fix numbers $E\in\DR$, $m>0$ and suppose that all balls $\ball_\ell(x)$
inside $\ball_L(u)$ are \EmNS.
Then $\forall$ $y\in\cG\setminus\ball_L(u)$ the function
$$
f: x \mapsto |G_{\cG}(x,y;E)|
$$
is $(\ell,q)$-subharmonic in $\ball_L(u)$ with $q = \eul^{-\gamma(m,\ell)\ell}$.
\end{lemma}

\proof The claim follows directly from the Definition \ref{def:subh}.
\qedhere

\vskip 2mm

Lemma \ref{lem:subh.1} suffices to assess the Green functions in a ball $\ball_L(u)$ which does not contain any singular $\ell$-ball, but to analyze the situation where $\ball_L(u)$ contains
one singular ball $\ball_\ell(w)$ (more precisely, it does not contain any pair of disjoint
singular $\ell$-balls), one needs the following extension of Lemma \ref{lem:subh.1},
which exploits the idea used of the proof of Theorem 1 in \cite{Sp88}: approaching a single "bad" ball separately from the points $x$ and $y$.

\begin{lemma}\label{lem:subh.2}
Let $\cG$ be a finite connected graph, and $f:\cG\times\cG\to\DR_+$, $f:(x,y)\mapsto f(x,y)$, be a function which is separately $(\ell,q)$-subharmonic in $x\in\ball_{r'}(u')\subset \cG$ and in
$y\in\ball_{r''}(u'')\subset \cG$, with $r',r''\ge \ell\ge 0$ and $\rd(u', u'')\ge r'+r''+2$. Then
\be\label{eq:lem.subh.2}
f(u',u'') \le q^{ \left\lfloor \frac{r'+1}{\ell+1} \right\rfloor
                 + \left\lfloor \frac{r''+1}{\ell+1} \right\rfloor } \cM(f,\cG\times\cG)
           \le q^{ \frac{r'+r'' - 2\ell}{\ell+1} } \cM(f,\cG\times\cG).
\ee
\end{lemma}

\proof
For each $y''\in\ball_{r''+1}(u'') $ define the function $f_{y''}:x'\mapsto f(x',y'')$
in $\cG$. By assumption, it is $(\ell,q)$-subharmonic in
$\ball_{r'}(u')$, so Lemma \ref{lem:subh.1} implies,
$$
\forall\, y''\in\ball_{r''+1}(u'') \quad
f(u',y'') = f_{y''}(u') \le q^{ \frac{r'-\ell}{\ell+1}} \cM(f,\cG\times\cG).
$$
Consider now another function, $\tf_{u'}: y''\mapsto f(u', y'')$, $y''\in\cG$.
It is $(\ell,q)$-subharmonic in $\ball_{r''}(u'')$, by hypothesis. The above inequality
reads as
$$
\cM(\tf_{u'}, \ball_{r''+1}(u'')) \le q^{ \frac{r'-\ell}{\ell+1}} \cM(f,\cG\times\cG),
$$
so another application of Lemma \ref{lem:subh.1} proves the claim:
$$
f(u',u'') = \tf_{u'}(u'') \le q^{ \frac{r'-\ell}{\ell+1}} \cM(\tf_{u'}, \ball_{r''+1}(u''))
\le q^{ \frac{r' + r'' -2\ell}{\ell+1}} \cM(f,\cG\times\cG).
\qedhere
$$


\section{Fixed-energy scale induction}
\label{sec:fixed.energy}

\subsection{Scaling of Green functions in absence of tunneling}

\begin{definition}
A ball $\ball_{L_{k+1}}(u)$ is called $E$-tunneling (\ET) if it contains two disjoint
\EmS balls of radius $L_k$, and $E$-non-tunneling (\ENT), otherwise.
\end{definition}

\begin{lemma}\label{lem:NR.NT.implies.NS}
If a ball $\ball_{L_{k+1}}(u)$ is \ENR and \ENT, then it is \EmNS.
\end{lemma}

\begin{center}
\begin{figure}\label{fig:f1}
\begin{tikzpicture}
\clip (0,0) rectangle ++(12,8);

\draw (3, 4) circle (1.85);
\draw (3, 4) circle (0.27);
\fill (3,4) circle (0.05);
\node at (3,4) (xc) {};
\node at (2.85,4.4) (x) {$x$};
\node at (2.0,3.2) (ballx) {$\ball_{L_k}(x)$};
\node at (1.7, 7.0) (ballrx) {$\ball_{r'}(x)$};
\node at (2.9, 5.95) (pballrx) {};
\draw[->] (1.75, 6.75) to [bend left]   (pballrx.north) ;

\node at (2.8, 4) (pballx) {};
\draw[->] (2.1, 3.5) to [bend left]   (pballx.west) ;

\draw[->] (3,4) -- ++(1.63, 0.60) ;
\node at (3.8, 4.6) (rx) {$r'$};

\filldraw[color=black, fill=lightgray!40!white] (5.47, 4.9) circle (0.70);
\fill (5.47, 4.9) circle (0.05);
\node at (5.47, 4.9) (wc) {};
\node at (5.5, 5.12) (w) {$w$};

\node at (4.5, 6.7) (ballw) {$\ball_{2L_k}(w)$};
\node at (5.30, 5.6) (pballw) {};
\draw[->] (4.5, 6.3) to [bend left]   (pballw.north) ;

\draw (9, 4) circle (2.85);
\draw (9, 4) circle (0.27);
\fill (9,4) circle (0.05);
\node at (9,4) (yc) {};
\node at (9.25,4.39) (y) {$y$};
\draw[->] (9,4) -- ++(-2.7, 0.77) ;
\node at (7.75, 4.7) (ry) {$r''$};

\node at (10.0,2.2) (bally) {$\ball_{L_k}(y)$};
\node at (9.2, 3.8) (pbally) {};
\draw[->] (10, 2.5) to [bend right]   (pbally.east) ;

\node at (6.0, 7.5) (ballry) {$\ball_{r''}(y)$};
\node at (7.3, 6.45) (pballry) {};
\draw[->] (6.3, 7.2) to [bend left]   (pballry.north) ;

\draw (7.47, 2.9) circle (0.27);
\fill (7.47, 2.9) circle (0.03);

\draw (3.47, 2.9) circle (0.27);
\fill (3.47, 2.9) circle (0.03);
\node at (3.5, 1.2) (ballsEmNS) {};
\node at (3.0, 1.0) (ballv) {$(E,m)$-NS balls $\ball_{L_k}(v)$};
\node at (3.5,2.6) (pbvx) {};
\draw[->] (ballsEmNS) to [bend right]   (pbvx.east) ;

\node at (7.2, 2.85) (pbvy) {};
\draw[->] (ballsEmNS) to [bend left]   (pbvy.west) ;

\end{tikzpicture}
\caption{Example for the proof of Lemma \ref{lem:NR.NT.implies.NS}.}
\end{figure}
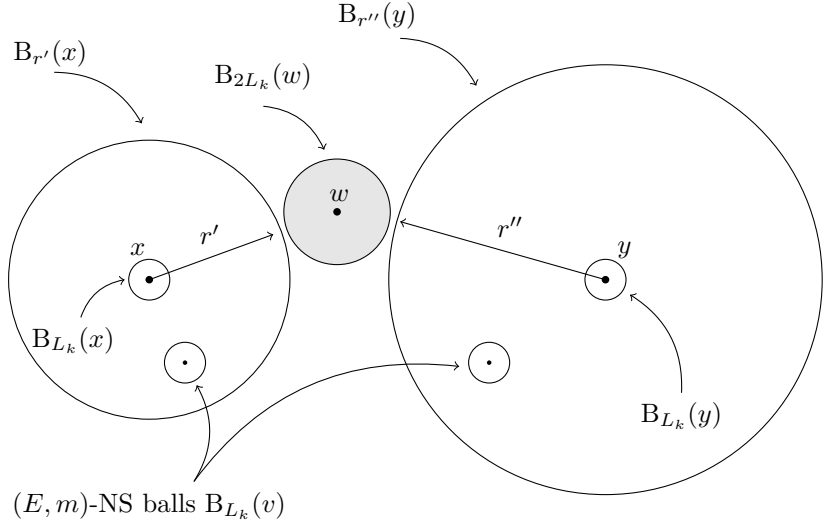

\end{center}

\proof (See Fig. 1.)
Fix two points $x,y\in\ball_{L_{k+1}}(u)$ with
$\rd(x,y) \ge L_{k}^{1+\varrho} = L_{k}^{7/6}$ and let
$R = \rd(x,y)$, so $R - 2L_k \ge R(1 - 2 L_k^{-\varrho})$.
Since $\ball_{L_{k+1}}(u)$ is \ENT, there is a ball $\ball_{2L_k}(w)$
such that any $L_k$-ball disjoint with $\ball_{2L_k}(w)$ is
\EmNS. By triangle inequality, there are integers $r', r''\ge 0$ such that
$r'+r''\ge R - 2L_k$, the balls $\ball_{r'}(x)$ and $\ball_{r''}(y)$ are disjoint
from each other and from $\ball_{2L_k}(w)$, so
any ball $\ball_{L_k}(v)$ inside $\ball_{r'}(x)$ and inside $\ball_{r''}(y)$ is
\EmNS.

Assume first that $r'\ge L_k$ and $r''\ge L_k$ (otherwise,
one of the points $x,y$ is covered by $\ball_{2L_k}(w)$, so
one of the radii $r', r'' \ge R - 3L_k$ and the same argument as below applies).
By Lemma \ref{lem:GF.is.SH.1}, the function
$f:(v,z)\mapsto | G_{\ball_{L_{k+1}}}(v,z;E)|$
is $(L_k, q)$-subharmonic in
$v\in\ball_{r'}(x)$ and in $z\in\ball_{r''}(y)$, with
$q\le \eul^{-mR(1 + L_k^{-\tau}) }$.
By Lemma \ref{lem:subh.2},
one can write, with the convention $-\ln 0 = +\infty$:
\begin{align*}\label{eq:proof.NS}
-\ln f(x,y) &\ge
-\ln
\Bigg\{ \left(
e^{ -m(1+ L_{k}^{-\tau})L_{k} }
\right)^{\frac{R\left( 1 - 2 L_k^{- \varrho} \right)}{L_{k}+1}}
e^{ L_k^\beta} \Bigg\}
\\
& \ge mR\left( \left(1+ {\textstyle \half } L_{k}^{-\tau} \right)
    {\textstyle \frac{L_{k}}{ L_{k} + 1} } \left( 1 - 2 L_k^{- \varrho} \right)
    -  L_k^{1 - \beta} m^{-1} R^{-1} \,\right)
\intertext{ (with $m\ge 1$,
$R \ge L_k^{1+\varrho} = \rd(x,y)\ge L_k^{7/6}$\, $\beta=1/2$ and $L_0$ large enough)}
 &\ge
 mR \left( (1 + L_{k}^{-1/8})  \big(1 - 3 L_k^{-1/6} \big)
    -  L_k^{-2/3} \right)
\\
& \ge  m \rd(x,y) \left(1 +{\textstyle \half } L_{k}^{-1/8} \right)
\\
&\ge \gamma(m,L_{k+1}) \,\rd(x,y)  + \ln |\pt \ball_{L_{k+1}}(u)|,
\end{align*}
as required for the \EmNS property of the ball $ \ball_{L_{k+1}}(u)$.
\qedhere

\subsection{Scale induction}

Introduce the following notations:
$$
\bal
P_k &= \sup_{x\in\cZ} \pr{ \ball_{L_k}(x) \text{ is \EmS}},
\\
Q_k &= 2 \sup_{x\in\cZ} \pr{ \ball_{L_k}(x) \text{ is \ER}}
\le 2C_W\, C_d L_k^d\, \eul^{ -L_k^{\beta}}
\eal
$$
(the latter inequality uses the Wegner estimate  \eqref{eq:Wegner}).

\begin{theorem}\label{thm:MSA.main.fixed}
If there is an integer $L_0\ge 1$ such that \eqref{eq:Lipshitz} is fulfilled and
$$
\textstyle
\min \{P_0, Q_0\} \le C_d^{-2} L_{0}^{-\kappa}, \; \kappa\ge \frac{2\alpha d}{2 - \alpha},
$$
then for all $k\ge 0$, $P_k \le C_d^{-2}\, L_{k}^{-\kappa}$.
\end{theorem}

\proof
$\,$
(The main argument combines the ideas from \cite{Sp88} and \cite{GK01}.)
By virtue of Lemma \ref{lem:NR.NT.implies.NS}, if a ball $\ball_{L_{k+1}}(u)$ is \EmS,
then it is either \ER or \ET. There are $<\half C^2_d L_{k+1}^{2d}$
pairs of disjoint  $L_k$-balls in $\ball_{L_{k+1}}(u)$, thus
$$
P_{k+1} \le \half C_d^2 L_{k+1}^{2d} \cP_k^2 + \half Q_{k+1}.
$$
By Wegner estimate \eqref{eq:Wegner},
$Q_{k+1} \le C_W C_d L_{k+1}^d  \eul^{-L_{k+1}^{1/2}}$.
An elementary calculation shows that the function
$$
\bal
f: L &\mapsto \ln \left( \const L^{-\kappa}\right)
-\ln \left(\const\, L^d  \eul^{-L^{1/2}} \right)
%
= L^{1/2} - \const \ln L
\eal
$$
on $[1, +\infty)$ is either non-negative or admits a unique zero. In either case, the assumption
$Q_0 \le C_d^{-2} L_{0}^{-\kappa}$ implies $Q_{k+1} \le C_d^{-2} L_{k+1}^{-\kappa}$
for all $k\ge 0$.
Thus by induction on $k=0, 1, \ldots$,
$$
\bal
P_{k+1} &\le 2 \cdot \half  C_d^2 L_{k+1}^{ 2d } P_k^{2}
%
\le \frac{C_d^2}{\big(C_d^{2}\big)^2} L_{k+1}^{-\frac{2\kappa}{\alpha} + 2d}
\le C_d^{-2} L_{k+1}^{-\kappa}.
\eal
$$
provided that $\frac{2\kappa}{\alpha} + 2d \ge \kappa$, i.e., for all
$\kappa\ge \frac{2\alpha d}{2 - \alpha}$.
\qedhere

\smallskip

This calculation shows that the power-law bound on the probabilities $\cP_k$ is reproduced
at each scale $L_k$ with the same exponent $\kappa>0$. Actually, it comes with a small bonus which seems to have never been used, until recently. Pick a value
$\kappa > \frac{2\alpha d}{2 - \alpha}$, so that
$\theta :=\left(\frac{2}{\alpha} - \frac{2d}{\kappa}\right) - 1 >0$, and observe that
$$
\textstyle{
\frac{2\kappa(1+\theta)^k}{\alpha} - 2d
\ge \frac{2\kappa(1+\theta)^k}{\alpha} - 2d
= \kappa (1+\theta)^k \left\{ \frac{2}{\alpha} - \frac{2d}{\kappa} \right\}
=\kappa (1+\theta)^{k+1}.
}
$$
If for some $k_0\ge 0$ one has $P_{k_0} \le C_d^{-2} L_{k_0}^{-\kappa(1+\theta)^{k_0}}$, then by induction,
for all $k\ge k_0$,
$$
\bal
P_{k+1} &\le 2 \cdot \half  C_d L_{k+1}^{ 2d } P_k^{2}
%
\le \frac{C_d^2}{C_d^4} L_{k+1}^{-\frac{2\kappa(1+\theta)^k}{\alpha} + 2d}
\\
&\le  C_d^{-2} L_{k+1}^{-\kappa(1+\theta)^{k+1}}.
\eal
$$
This gives rise to the probabilities $P_k$ decaying \emph{faster that any power law}.

Observe also that taking $\alpha\downarrow 1$ results in $2\alpha/(2-\alpha)\downarrow 2$,
so that the exponent $\kappa>0$ in the \emph{hypothesis} of Lemma \ref{thm:MSA.main.fixed}
can be chosen arbitrarily close to $2d$.

\vskip4mm
\emph{We see that the complete, fixed-energy MSA procedure can be effectively reduced to
Lemma \ref{lem:NR.NT.implies.NS} and Theorem \ref{thm:MSA.main.fixed} and results in upper bounds
on the probability of singular balls decaying faster than any power-law.}

\section{From fixed to variable energy: First approach}
\label{sec:SW.ETV}

Now we establish a fairly general relation between fixed-energy
probabilistic estimates on the Green functions and variable-energy bounds
for two disjoint finite volumes. It does not matter how the probabilistic input is obtained; in particular, the results of this section can be combined both with the MSA,
performed for each fixed
energy $E$ in a given interval $I\subset \DR$, and with the FMM (which always starts
as a fixed-energy analysis).

It is convenient to assume that $|I|=1$, so the interval $I$ with the Lebesgue measure
$\mes(\cdot)$ is a probability space, and so is the product space $(\Om\times I, \DP\times \mes)$.
(The idea of using the "disorder-energy" space with product measure has been used earlier, e.g.,
in \cite{MS85} and \cite{SW86}.)
Given $L\in\DN$ and points $x,y\in\cZ$, set for brevity
\be\label{eq:def.Mxy}
\cM_{x,y}(E)= |G_{\ball_L}(x,y;E)|,
\;
\cM_{x}(E)=\max_{y\in\pt^- \ball_L(x)} \cM_{x,y}(E),
\ee
and introduce the subsets of $I$ parameterized by $a>0$:
\be
\cE_{x,y}(a) = \{E\in I:\, \cM_{x,y}(E) \ge a \},\;\;
\cE_{x}(a) = \{E\in I:\, \cM_{x}(E) \ge a \}.
\ee
(The $L$-dependence will be often omitted for brevity.)


\begin{theorem}\label{thm:sing.width}
Let $L\ge 0$, $x\in\cZ$, $y\in\pt^-\ball_L(x)$. Let $\{\lam_j\}_{j=1}^{N}$  be the eigenvalues
of the operator $H_{\ball_L(u)}(\om)$ and $I\subset\DR$ an interval of unit length.
Let be given numbers $a,b,c, \cP_L>0$
such that
\be\label{eq:cond.a.b.c}
b \le \min\left\{ |\ball_L(u)|^{-1} ac^{2}, \, c \right\},
\ee
and for all $E\in I$
\be\label{eq:thm.fixed.E.intervals}
\pr{ \cE_{x}(a)} \equiv \pr{ \cM_{x}(E) \ge a } \le \cP_L.
\ee
There is an event
$\cB_{x}(b)$ with $\pr{\cB_{x}(b)}\le b^{-1} \cP_L$ such that
$\forall$ $\om\not\in\cB_{x}(b)$, the  set
$$
\cE_{x}(2a) = \cE_{x}(2a;\om) = \Big\{ E: \, \cM_{x}(E) \ge 2 a \Big\}
$$
is contained in a union of intervals
$\cup_{j=1}^N I_j$,
$I_j := \{E: |E-\lam_j| \le c \}$, $\lam_j\in I$.
\end{theorem}
\proof
Consider the following events parameterized by $b>0$:
\be\label{def:cS.cE.cB}
\bal
\cB_{x}(b)
&= \myset{ \om\in \Om:\, \mes( \cE_{x}(a) ) > b }.
\eal
\ee
Apply Chebyshev's inequality and the Fubini theorem
combined with \eqref{eq:thm.fixed.E.intervals}:
\begin{align}
\pr{ \cB_{x}(b) }
&\le b^{-1} \esm{ \mes( \cE_{x}(a) )  }
\nonumber
\\
& = b^{-1} \int_I \, dE\,\, \esm{ \one_{\{\cM_{x}(E) \ge a \}} }
\label{eq:prob.cB}
 \le b^{-1} \cP(L).
\end{align}
Now fix any $\om\not\in \cB_{x}(b)$, so that
$
\mes( \cE_{x}(a); \om ) \le b.
$
There is a subset $\{\lam_j \}_{j=1}^{N'}$ of the EVs of the operator $H_{\ball_L(x)}$ such that
the Green function $E\mapsto G_{\ball_L(x)}(x,y;E)$  reads as a rational function
(below we remove the vanishing terms, if any)
\be\label{eq:GF.rational}
f: E \mapsto G_{\ball_L(x)}(x,y;E) =: \sum_{j=1}^{N'} \frac{\kappa_j}{\lam_j - E},
\;\; N' \le N := |\ball_L(x)|;
\ee
here $\kappa_j = \kappa_j(x,y)\ne 0$ and
$\sum_j |\kappa_j| \le \sum_i |\psi_i(x)\psi_i(y)| \le N$. Let
$$
\bal
\cR(2c) &= \big\{ \lam\in\DR:\;  \min_j \, |\lam_j - \lam| \ge 2 c \big\},
\\
\cR(c) &= \big\{ \lam\in\DR:\;  \min_j \, |\lam_j - \lam| \ge  c \big\},
\quad c>0.
\eal
$$
Observe that, with $0<b\le c$,  $\cA_b:=\{E:\, \dist(E, \cR(2c)) < b \}\subset\cR(c)$,
hence, the set $\cA_b$ is a union of open sub-intervals at distance $\ge c$ from the spectrum,
and on each
sub-interval one has $|f'(E)| \le N c^{-2}.
$
Let us show
by contraposition
that, with $\om\not\in \cB_{x,y}(b)$,
$$
\myset{E:\, |G_{\ball_L}(x,y;E)| \ge 2 a } \cap \cR(c)=\varnothing.
$$
Assume otherwise, pick any point $\lam^*$ in the non-empty set in the LHS,
and let $J := \{ E': \, |E' -\lam^*|\le b\}\subset\cA_b\subset\cR(c)$.
Then for any
$E\in J$
one has, by \eqref{eq:cond.a.b.c},
$$
\bal
|f(E)| & \ge |f(\lam^*)| - | J| \sup_{E'\in J} |f'(E')|
%
> 2 a - N c^{-2} \cdot b  \ge  a,
\eal
$$
so
$J \subset \cE_{x,y}(a)$  and
$\mes(\cE_{x,y}(a) ) \ge \mes(J) = 2 b > b$, contrary to the choice of $\om$.
Since the set $\cR(c)$ is independent of $y$, the assertion follows from \eqref{eq:prob.cB}.
\qedhere

\smallskip

Below we provide some possible choices of the parameters $a, b, c$ (depending, of course, upon the scale $L$) in three most frequently used frameworks.

\begin{enumerate}
  \item Weaker MSA-type bounds: $\cP_{L_k} = P_k \le L_k^{-\kappa(1+\theta)^k}$,
  $\kappa > \frac{\alpha d}{2-\alpha}$, $\alpha=3/2$. One can set,
  for $L \in\{L_k, k\ge 0\}$,
$$
a(L_k) = L_k^{-\frac{3\kappa}{5}(1+\theta)^k}, \;\;
b(L_k) = L_k^{-\frac{\kappa}{5}(1+\theta)^k}, \;\;
c(L_k) = L_k^{-\left( \frac{\kappa}{5} - \frac{d}{2} \right) (1+\theta)^k}
$$
  \item Sub-exponential MSA-type bounds: $\cP_{L_k} \le \eul^{-L_k^{\delta}}$, $\delta>0$. Then one can set
$$
a(L_k) =  e^{-\frac{1}{3}L_k^{\delta}}, \;\;
b(L_k) = e^{-\frac{2}{3}L_k^{\delta}}, \;\;
c(L_k) = \eul^{ - \frac{1}{8}L_k^{\delta}}.
$$
  \item FMM-type bounds: $\cP_{L} \le \eul^{-mL}$, $m>0$. Then one can set,
  for all $L\in\DN$ large enough,
$$
a(L) = e^{-\frac{1}{3}mL}, \;\; b(L) = e^{-\frac{2}{3}mL}, \;\;
c(L) = \eul^{ - \frac{m}{8}L}.
$$
\end{enumerate}

\begin{theorem}\label{thm:SW.2vol}
Assume the condition \eqref{eq:Lipshitz}. Suppose that for some $L\in\DN$,
numbers $a=a(L)$, $b=b(L)$, $c=c(L)$ and $\cP_L>0$ obey \eqref{eq:cond.a.b.c}, and for
some interval $I\subset\DR$ and
all $E\in I$, for any ball $\ball_L(x)\subset\cZ$
\be\label{eq:thm.fixed.E.intervals.2}
\pr{ \cM_{x}(E) \ge a } \le \cP_L.
\ee
Then for any two disjoint balls $\ball_L(x),\, \ball_L(y)\subset\cZ$ the following bound holds true:
\be
\pr{ \exists\, E\in I:\, \min(\cM_{x}(E), \cM_{y}(E)) > a(L)}
\le 4C_W C_d^2 L^{2d}\, c(L) + \frac{2\cP_L}{b(L)}.
\ee
\end{theorem}

\proof
Let the events $\cB_x(b)$, $\cB_y(b)$ be defined as in \eqref{def:cS.cE.cB} and
$\cB = \cB_x\cup\cB_y$, then
\be\label{eq:thm.SW.2vol.1}
\bal
&\pr{ \cE_{x}(a) \cap \cE_{y}(a) \ne \varnothing }
\le \pr{ \cB }  +
\esm{ \pr{ \left\{ \cE_{x}(a) \cap \cE_{y}(a) \ne \varnothing \right\} \cap \cB^\rc } }
\\
& \qquad\qquad\quad \le 2b^{-1}\cP_L +
\esm{\pr{ \left\{ \cE_{x}(a) \cap \cE_{y}(a) \ne \varnothing \right\} \cap \cB^\rc \,|\,
    \cF_{\boxx_L(y)}}}.
\eal
\ee
It remains to assess the conditional probability in the RHS. For $\om\not\in\cB^\rc$,
each of the sets $\cE_{x}(a)$, $\cE_{y}(a)$ is covered by intervals
of width $2c(L)$ around the respective EVs $\lam_i(x)\in\Sigma( H_{\ball_L(x)}$,
$\lam_j(y)\in\Sigma( H_{\ball_L(y)}$, and
for disjoint balls $\boxx_L(x), \boxx_L(y)$ these spectra are independent. Now apply Theorem \ref{thm:sing.width} and the Wegner estimate:
\be\label{eq:thm.SW.2vol.2}
\bal
\pr{ \left\{ \cE_{x}(a) \cap \cE_{y}(a) \ne \varnothing \right\} \cap \cB^\rc \,|\,
    \cF_{\boxx_L(y)}}
%
&
  \le |\ball_L(y)| \, \sup_{\lam\in I}
      \pr{ \dist( \cE_{x}(a), \lam) \le c }
\\
&
  \le |\ball_L(y)| \, \sup_{\lam\in I}
      \pr{ \dist( \Sigma( H_{\ball_L(x)}, \lam) \le 2c }
\\
&
  \le 4C_W |\ball_L(y)| \cdot |\ball_L(x)| \cdot  c(L).
\eal
\ee
Collecting \eqref{eq:thm.SW.2vol.1} and \eqref{eq:thm.SW.2vol.2}, the assertion follows.
\qedhere

\vskip4mm

It is clear that the above approach, albeit very general and based on a Wegner-type
bound, gives rise to exponential decay bounds on the Green functions
only if the fixed-energy analysis provides exponential probabilistic bounds; as it is
well-known, this has been achieved so far only in the framework of the FMM.

\subsection{Spectral localization}

The assertion of Theorem \ref{thm:2vol.VEMSA.PCT} has a structure
similar to that of the MSA bound from the work by von Dreifus and Klein \cite{DK89}.
More precisely, it guarantees a decay rate of Green functions slower than exponential,
but faster than any power-law. It is not difficult to adapt the well-known argument from
\cite{DK89} and prove that with probability one,
all polynomially bounded solutions to the equation $H(\om) \psi = E\psi$ are in fact
square-summable. The latter property requires a Shnol--Simon type result
on spectrally a.e. polynomial boundedness of generalized eigenfunctions.
It will follow independently by RAGE (Ruelle--Amrein--Georgescu--Enss) theorems
(see a detailed discussion along with a bibliography, e.g., in \cite{CyFKS87}) from
the dynamical localization proven in Section \ref{sec:MSA.to.DL}.

\vskip4mm
\section{From fixed to variable energy: Second approach}
\label{sec:SW.PCT}

\subsection{The spectral reduction}

\begin{theorem}\label{thm:SW.CPT}
Let be given a ball $\ball_L(x)$, $L\ge 1$, and
numbers $a(L)$, $b(L)$, $c(L)$, $\cP_L>0$ obeying \eqref{eq:cond.a.b.c} and
such that, for some interval $I$, all $E\in I$,
\be\label{eq:thm.fixed.E.intervals.3}
\pr{ \cM_{x}(E) \ge a } \le \cP_L.
\ee
Then the following properties hold true:
\begin{enumerate}[\rm(A)]
\item
For any $b\ge \cP_L$ there exists an event $\cB_b$
such that $\pr{\cB_b}\le b^{-1}\cP_L$
and for any $\om\not\in\cB_p$ the set of energies
$$
\cE_{x}(a) = \cE_{x}(a;\om) = \{ \cM_{x}(E) \ge a\} \cap I
$$
is covered by $K < 3N^2$ intervals $J_i = [E^-_i,E^+_i]$ of total length
$\sum_i |J_i|\le b$.
  \item The endpoints $E^\pm_i$ are determined by the function
  $E\mapsto \langle \one_x\,|\, (H_{\bball_L(u)} - E)^{-1} \,|\, \one_y \rangle$
  in such a way that, for the one-parameter
  family $A(t) := H_{\ball_L(u)} + t\one$, the endpoints $E^\pm_i(t)$ for
  the operators $A(t)$ (replacing operators $H_{\ball_L(u)}$) have the form
$$
   E^\pm_i (t) = E^\pm_i (0) + t, \quad t\in\DR.
$$

\end{enumerate}

\end{theorem}

\proof
(A) Fix a point $y\in\pt^-\ball_L(x)$ and consider the rational function
$$
f_y: E \mapsto \sum_{i=1}^N \frac{\kappa_i}{ \lam_i - E}
:= \sum_{i=1}^N \frac{\psi_i(x) \, \psi_i(y)}{ \lam_i - E}.
$$
Its derivative has the form
$$
f'_y(E)  = \sum_{i=1}^N \frac{-\kappa_i}{ (\lam_i - E)^2}
=: \frac{\csP(E)}{ \csQ(E)}, \;\; \deg \, \csP \le 2N-2,
$$
and has $\le 2N-2$ zeros and $\le N$ poles, so $f_y$ has $\le 3N-1$ intervals of monotonicity
$I_{i,y}$, and the total number of monotonicity intervals of all functions
$\{f_y, y\in\pt^-\ball_L(x)\}$ is bounded by
$K\le|\pt^-\ball_L(x)|(3N-1)\le|\ball_L(x)|(3N-1)<3N^2$, so
$$
\cup_{y\in\pt^-\ball_L(x)}
\{E:\; f_y(E) \ge a\} = \cup_{i=1}^{K} J_i, \quad
J_i = [E^-_i,E^+_i] \subset I,
$$
where, obviously, $\sum_i |J_i|\le\mes\, \{E:\; \cM_x(E) \ge a\}$.
\par\vskip2mm
\noindent
(B)  Consider a one-parameter operator family
$
A(t) = H_{\ball_L(u)}(\om)  + t\one.
$
All these operators share common
eigenvectors; the latter determine the coefficients $\kappa_i$, so one can choose
eigenfunctions $\psi_i(t)$ constant in $t$ and obtain $\kappa_i(t) = \kappa_i(0)$.
The eigenvalues of operators $A(t)$ have the form $\lam_i(t) = \lam_i(0)+t$. We conclude
that the Green functions, with fixed $x$ and $y$, have the form
$f_{x,y}(E;t) = f_{x,y}(E-t;0)$, so that the intervals $J_i(t)$ have indeed the form
$J_i(t) = [E^-_i + t, E^+_i + t]$.
\qedhere

\begin{theorem}\label{thm:SW.2vol.PCT}
Consider two disjoint balls $\ball_L(x)$, $\ball_L(y)$ and
the random variables
$$
\xi_x(\om) := |\ball_L(x)|^{-1}\sum_{z\in\ball_L(x)} V(z;\om),\quad
\eta_z(\om) := V(z;\om) - \xi_x(\om), \; z\in\ball_L(x)
$$
(the sample average and fluctuations of the potential in $\ball_L(x)$).
Let $\fF_{x}$ be the sigma-algebra generated by the random variables
$\{\eta_y, y\in\ball_L(y); V(z;\cdot), z\not\in\ball_L(x)\}$.
Consider the conditional probability distribution function
$$
F_{\xi_x}(t\,|\, \fF_x) = \pr{ \xi_x \le t \,|\, \fF_x}
$$
and its continuity modulus
$$
\nu_{\xi_x}( s\,|\, \fF_x) =
\sup_{t\in\DR}\; \essup  \;
\left( F_{\xi_x}(t+s\,|\, \fF_x) - F_{\xi_x}(t\,|\, \fF_x)\right).
$$
Suppose that, for some $C,C', A,A',B,B'\in(0,+\infty)$
\be\label{eq:cond.PCT}
\forall\, s\in[0,1] \qquad
\pr{ \nu_{\xi_x}( s\,|\, \fF_x) > C L^{A} s^B }\le C' L^{A'} s^{B'}.
\ee
Then
\be
\pr{\exists\, E\in I:\; \min\{ \cM_x(E), \cM_y(E)\} \ge a} \le N^2 \htil(4b)
\ee
where
$$
\htil(s) := C L^{A} s^{B} + C' L^{A'} s^{B'}.
$$
\end{theorem}

\proof

Using the decomposition $V(z;\om) = \xi_x(\om)\one + \eta_z(\om)$
in the ball $\ball_L(x)$, , consider
the respective operator decomposition
$$
H_{\ball_L(x)}(\om) = A_x(\om) + \xi_x(\om)\one,
\quad A(\om) = H_0 + \eta_\bullet(\om),
$$
where, conditional on $\fF_x$, the operator $A(\om)$ is non-random. Fix
a number $b>0$ and consider the events $\cB_b(x)$ (relative to the operator
$H_{\ball_L(x)}$) and, respectively, $\cB_b(y)$; let $\cB = \cB_b(x)\cup\cB_b(y)$.
For any $\om\not\in\cB$, the energies $E$ where $\cM_x(E)\ge a$
are covered by intervals $J_{i,x} = [E^-_{i,x}, E^+_{i,x}]$,
with $\sum_i |J_{i,x}|\le b$, and, respectively,
the energies $E$ where $\cM_y(E)\ge a$ are covered by intervals
$J_{i,y} = [E^-_{i,y},E^+_{i,y}]$, also obeying
$\sum_i |J_{i,y}|\le b$. Conditional on $\fF_x$, intervals
$J_{i,y}$ become non-random, while for the intervals $J_{i,x}$
we can write, by virtue of assertion (B) of Theorem \ref{thm:SW.CPT},
$$
J_{i,y}(\om) =
[E^{(-,0)}_{i,x}(\om) + \xi_x(\om), \; E^{(+,0)}_{i,x}(\om) + \xi_x(\om)]
$$
where $E^{(\pm,0)}_{i,x}(\om)$ are $\fF_x$-measurable, i.e., non-random under the conditioning
by $\fF_x$.

Further, set $\eps_{i,x} = |J_{i,x}|$ and $\eps_{j,y} = |J_{j,y}|$, then
$$
\bal
\{\om:\, J_{i,x}\cap J_{j,y} \ne \varnothing\}
&\subset
\left\{ \big| E^{(-,0)}_{i,x}(\om) - E^{(-,0)}_{j,y}(\om) \big| \le \eps_{i,x} + \eps_{j,y}
\right\}
\\
&= \left\{ \big| \xi_x(\om) - \tE(\om) \big| \le \eps_{i,x} + \eps_{j,y}
\right\}
\eal
$$
with $\tE(\om)$ non-random under the conditioning. Apply the assumption \eqref{eq:cond.PCT}:
$$
\bal
& \pr{ \big| \xi_x(\om) - \tE(\om) \big| \le \eps_{i,x} + \eps_{j,y} }
\le \pr{ \big| \xi_x(\om) - \tE(\om) \big| \le 4b }
\\
& \qquad \le \pr{ \nu_{\xi_x}( 4b\,|\, \fF_x) > C L^{A} (4b)^B }
+ C L^{A} (4b)^{B}
\\
& \qquad = \htil(4b).
\eal
$$
Taking the sum over $i, j$,  we obtain the erquired bound:
$$
\bal
\pr{\exists\, E\in I:\; \min\{ \cM_x(E), \cM_y(E)\} \ge a }
%
&\le \sum_{i,j} \pr{\om:\, J_{i,x}\cap J_{j,y} \ne \varnothing}
\\
&  \le N^2 \htil(4b).
\qedhere
\eal
$$
\par
\vskip4mm

In particular, taking into account Theorem \ref{thm:MSA.main.fixed}, we can set, for $L=L_k$,
$$
a=a(L_k)=\eul^{-\gamma(m,L_k)L_k}, \;
\cP_{L_k} = L_k^{-\kappa(1+\theta)^k}, \;
b=b(L_k) =L_k^{-\frac{\kappa}{2}(1+\theta)^k}.
$$
These settings give rise to the following corollary of Theorem \ref{thm:SW.2vol.PCT}:

\begin{theorem}\label{thm:2vol.VEMSA.PCT}
If there is an integer $L_0\ge 1$ and numbers $m\ge 1$, $\alpha\in(1,2)$ such that \eqref{eq:Lipshitz} is fulfilled and
$$
\textstyle
\min \{P_0, Q_0\} \le C_d^{-2} L_{0}^{-\kappa}, \; \kappa > \frac{2\alpha d}{2 - \alpha},
$$
then for some $\theta>0$ and all $k\ge 0$, for any interval $I\subset\DR$ with $|I|\le 1$,
$$
\pr{E\in I:\; \ball_L(x) \text{ and $\ball_L(y)$ are \EmS} }
\le L_k^{-\frac{\kappa}{2}(1+\theta)^k}.
$$
\end{theorem}

\subsection{On the validity of the assumption \eqref{eq:cond.PCT}}

First of all, recall that, by an elementary result on Gaussian distributions, if  $V:\cZ\times\Om\to\DR$ is an IID Gaussian field, say, with zero mean and unit variance,
the sample average $\xi_x$ of the sample $\{V(z;\om), z\in\ball_L(x)\}$ is independent of
the sigma-algebra generated by the "fluctuations" $\eta_z(\om)$; moreover, it has Gaussian distribution $\cN(0, |\ball_L(x)|)$ and admits a probability density
with $\|p_{\xi_x}\|_\infty\le \frac{1}{\sqrt{2\pi}}\,|\ball_L(x)|^{1/2}$. In this particular case,
Eqn \eqref{eq:cond.PCT} can be replaced by a stronger, deterministic bound: the conditional
continuity modulus $\nu_{\xi_x}( s\,|\, \fF_x)$ is actually independent of the condition and is
bounded by $\|p_{\xi_x}\|_\infty \cdot s$.

Such a situation is rather exceptional, as shows the example of two IID random variables
$V_1(\om), V_2(\om)$ with uniform distribution ${\rm Unif}([0,1])$. Indeed, in this case
$\xi := (V_1+V_2)/2$, $\eta = (V_1-V_2)/2$ and the distribution of
$\xi$ conditional on $\eta$ is uniform on the interval $I_\eta$ of length $O(1-|\eta|)$, hence,
with constant density $O(\big|1 - |\eta|\big|^{-1})$, for $|\eta|<1$; for $\eta = \pm 1$,
this distribution is concentrated on a single point. However, this example shows also
how such a difficulty can be bypassed: excessively "singular" conditional distrubutions
of the sample mean $\xi$ occur only for a set of conditions having a small probability.
Using this simple idea, Gaume \cite{G10}, in the framework of his PhD project,
established the property \eqref{eq:cond.PCT} for
IID random fields with piecewise constant marginal \emph{probability density}. By standard
approximation arguments, it can be easily extended to piecewise Lipshitz (or H\"{o}lder)
continuous \emph{densities}, which is sufficient for
most physically relevant applications.
We believe that some variant of the property \eqref{eq:cond.PCT},
perhaps weaker but still sufficient for the purposes of the MSA, holds true
in a larger class of IID random fields.

\subsection{Exponential spectral localization}

The assertion of Theorem \ref{thm:2vol.VEMSA.PCT} has the same form as in the
conventional MSA bound going back to the work by von Dreifus and Klein \cite{DK89}
(actually, even slightly stronger); therefore, the same argument as in
\cite{DK89} (having its roots in \cite{FMSS85}) applies and proves that with probability one,
all polynomially bounded solutions to the equation $H(\om) \psi = E\psi$ are in fact
decaying exponentially fast at infinity, thus the operator $H(\om)$
has a.s. pure point spectrum. The latter property requires a Shnol--Simon type result
on spectrally a.e. polynomial boundedness of generalized eigenfunctions;
it will follow by RAGE theorems from
the dynamical localization proven in Section \ref{sec:MSA.to.DL}.

\section{From MSA to strong dynamical localization}
\label{sec:MSA.to.DL}

The first rigorous derivations of the dynamical localization from MSA-type probabilistic bounds on
the Green functions have been obtained by Germinet--De Bi\`{e}vre \cite{GD98} and
Damanik--Stollmann \cite{DS01}. The latter paper had a very eloquent title: "\emph{Multiscale analysis
implies strong dynamical localization}".  One of the main ingredient of these two works is the analysis of the so-called centers of localization
of square-summable eigenfunctions; this notion appeared earlier in the work \cite{DJLS96} which proved instrumental for a number of subsequent researches.
%
Later, Germinet and Klein \cite{GK01} discovered a substantially shorter argument, using
more efficiently  Hilbert--Schmidt norm estimates for spectral projections in a infinitely
extended configuration space. Formally,
\cite{GK01} considers operators in a Euclidean space $\DR^d$, but an adaptation to
a finite-dimensional lattice or, more generally, to a countable graph with polynomially
growing balls, is quite straightforward.

In the present paper, working with finite graphs, we bypass the `hard' analysis
of spectral projections and replace it by Bessel's  inequality.

The main result of this section can be summarized in the following meta-theorem,
expressing the `soft' argument by Germinet--Klein (viz., the finite-volume version thereof):
"\emph{The MSA bounds are essentially equivalent to the strong dynamical localization}",
with a meta-proof: "\emph{Apply Bessel's inequality}".

Owing to the results of Sections \ref{sec:SW.ETV}--\ref{sec:SW.PCT},
it actually suffices to
perform only the fixed-energy MSA, even in its simplest form proposed  in
\cite{Sp88}.

The extension to an infinite configuration space also admits a short meta-proof,
going back to the works by Aizenman \emph{et al.}: "\emph{Apply Fatou lemma}".


\subsection{EF correlators in finite balls}
\label{sec:MSA.to.DL.finite}

Given an interval $I\subset\DR$, denote by $\csB_1(I)$ the set of all Borel functions $\phi:\DR\to\DC$ with
$\supp\,\phi\subset I$ and $\|\phi\|_\infty \le 1$.

\begin{theorem}\label{thm:GK}
 Fix an integer $L\in\DN^*$ and assume that  the following bound holds for any pair of disjoint balls $\ball_L(x), \ball_L(y)$ and some quantity $\zeta(L)>0$:
$$
\pr{ \exists\, E\in I:\, \text{ $\ball_L(x)$ and $\ball_L(y)$ are $(E,m)${\rm-S}} } \le \zeta(L).
$$
Then for any $x,y\in\cZ$ with $\rd(x,y)> 2L+1$, any finite connected subgraph (of $\cZ$)
$\cG\supset\ball_L(x) \cup \ball_L(y)$ and any Borel function $\phi\in\csB_1(I)$
\be\label{eq:thm.MSA.to.DL}
\esm{ \big|\langle\one_x | \phi(H_\cG(\om)) | \one_y \rangle \big| }
\le 4 \eul^{-mL} + \zeta(L).
\ee
\end{theorem}

\proof Fix points $x,y\in\cZ$ with $\rd(x,y)> 2L+1$ and a graph
$\cG\supset\ball_L(x) \cup \ball_L(y)$. The operator $H_\cG(\om)$
has a finite orthonormal eigenbasis $\{\psi_i\}$ with respective eigenvalues
$\{\lam_i\}$. Let $\BbS = \pt \ball_L(x) \cup \pt \ball_L(y)$
(recall: this is a set of \emph{pairs}
$(u,u')$); note that $|\BbS|\le 2C_d^2L^d$, by \eqref{eq:ball.growth}. Suppose that for some $\om$, for each $i$ there is $z_i\in \{x,y\}$ such that
$\ball_L(z_i)$ is $(\lam_i,m)$-NS; let $\{v_i \}= \{x,y\}\setminus \{z_i\}$.
Denote
$\mu_{x,y}(\phi) = \big|\langle\one_x | \phi(H_\cG(\om)) | \one_y \rangle \big|$,
with $\mu_{x,y}(\phi)\le 1$.
Then by the GRI for the eigenfunctions,
%
\begin{align*}
 \mu_{x,y}(\phi)
& \le \|\phi\|_\infty \, \sum_{\lam_i \in I} |\psi_i(x) \psi_i(y)|
\le \sum_{\lam_i \in I} |\psi_i(z_i) \psi_i(v_i)|
\\
& \le \sum_{\lam_i \in I} |\psi_i(v_i)| \, \eul^{-mL} (C_d^2L^d)^{-1}
      \sum_{(u,u')\in\pt \ball_L(z_i)} |\psi_i(u)|
\qquad\qquad\qquad\qquad\qquad\qquad
\\
& \le \eul^{-mL} \sum_{\lam_i \in I} \; (C_d^2L^d)^{-1} \sum_{(u,u')\in \BbS}
      |\psi_i(u)| \left(|\psi_i(x)| + |\psi_i(y)|  \right)
\\
&\le \eul^{-mL} \frac{|\BbS|}{ C_d^2L^d} \, \max_{u\in\cG} \sum_{\lam_i \in I}
    \half \left( |\psi_i(u)|^2 + |\psi_i(x)|^2 + |\psi_i(y)|^2 \right)
\\
\intertext{(using Bessel's inequality and $|\BbS|\le 2C_d^2L^d$)}
& \le  \eul^{-mL} \,2 \, \max_{u\in\cG}
       \left( 2\|\one_u\|^2 + \|\one_x\|^2 + \|\one_y\|^2 \right)
%
= 4 \eul^{-mL}.
\end{align*}
Denote
$\cS_L = \myset{\exists\, E\in I:\, \text{ $\ball_L(x)$ and $\ball_L(y)$ are $(E,m)$-S}}$,
with $\pr{\cS_L}\le \zeta(L)$, by assumption. Now we conclude:
$$
\esm{ \mu_{x,y}(\phi) } = \esm{ \one_{\cS_L}\mu_{x,y}(\phi) }
   + \esm{ \one_{\cS^{\rc}_L}\mu_{x,y}(\phi) }  \le \zeta(L) + 4\eul^{-mL}.
\qedhere
$$

It is clear that the exponential term $\eul^{-mL}$ can compete with $\zeta(L)$
only in applications to the FMM, for the MSA bounds on $\zeta(L)$ are at best sub-exponential
in $L$. Otherwise, $\zeta(L)$ is the dominant term.

Note also that the decay rate of the bound $\zeta(L)$ sets natural restrictions on the class of
the graphs $\cZ$, due to the presence of the `surface' factor $|\BbS|$,
$\BbS=\BbS(L)$. In particular,
only the FMM-based bounds have the chance to be efficient on trees and other graphs with exponentially growing balls.

\subsection{Dynamical localization on the entire graph}
\label{sec:MSA.to.DL.infinite}

Now one can make use of a simple argument employed earlier by Aizenman et al. \cite{A94,ASFH01},
in the framework of the FMM which always starts as a fixed-energy analysis. The quantities
$\mu_{x,y}^{(H)}(\phi) = \langle \one_x \,|\, \phi(H)\,|\, \one_y\rangle$ defined, for example, for bounded continuous or Borel functions $\phi$, generate signed (i.e., not necessarily positive)
spectral measures associated with a self-adjoint operator $H$:
$$
\int d\mu_{x,y}^{(H)}(E) \, \phi(E) := \langle \one_x | \phi(H)| \one_y\rangle.
$$
In particular, we can consider, with $x,y,u\in\cZ$ fixed, measures
$\mu^k_{x,y}$ related to operators $H_{\ball_{L_k}(u)}$, for all $k\ge 0$,
as well as their counterparts $\mu_{x,y}$ for the operator $H$
on the entire graph $\cZ$. A sufficient condition for the vague convergence
$\mu^k_{x,y}\rightarrow \mu_{x,y}$ as $k\to\infty$ is the strong resolvent convergence
$H_{\ball_{L_k}(u)} \rightarrow H$. Such convergence is well-known to occur for a very large class
of operators, including (unbounded) Schr\"{o}dinger operators in Euclidean spaces
and their analogs on the so-called quantum graphs. Indeed, for (not necessarily bounded)
operators $H_n$ with a common core $\cD$ to converge to an operator $H$ with the same core,
it suffices that $H_k \psi \to H\psi$ strongly for any element $\psi\in\cD$ (cf. \cite{Kato}).
For finite-volume operators, one can usually find an appropriate core $\cD$ formed by compactly supported functions $\psi$; for finite-difference Hamiltonians on graphs (even unbounded,
e.g., for DSO with unbounded potentials) one can choose
as $\cD$ the subset of all functions with finite supports. On such functions,
$H_{\ball_{L_k}(u)}\psi \to H\psi$ as $k\to\infty$ (by stabilization),
therefore, the spectral measures converge vaguely:
$\mu^k_{x,y}\rightarrow \mu_{x,y}$. By Fatou lemma,
for any bounded Borel set $A\subset\DR$,
one has
$$
\left| \mu_{x,y}(A) \right| \le \liminf_{k\to\infty} \left| \mu^k_{x,y}(A) \right|
$$
(here $\left| \mu(A) \right| := \sup \{ \mu(\phi), \, \|\phi\|\le 1, \, \supp\, \phi\subset A\}$).
Taking the expectation and using the uniform upper bounds on EF correlators in finite balls,
we conclude that
\be
\esm{ \sup_{\phi\in\csB_1} \, \big| \langle \one_x | \phi(H(\om)) | \one_y \rangle \big| }
\le C \eul^{ - a \ln^{1+c} \rd(x,y)}
\ee
(using the inequality
$L_k^{-\kappa(1+\theta)^k} \le C \eul^{ - a \ln^{1+c} L_k}$, $C,a,c>0$).
In particular, with functions $\phi_t:\lam\mapsto \eul^{-it\lam}$, we obtain the strong dynamical localization property for the ensemble of random Hamiltonians $H(\om)$.

\section{Sub-exponential bounds on EF correlators without bootstrap}
\label{sec:non.bootstrap}

Now we will show how the polynomial (or slightly stronger than polynomial) decay bounds
from Section \ref{sec:fixed.energy} can be substantially improved and replaced by sub-exponential
ones.
Germinet and Klein proved in \cite{GK01}
a highly optimized and very general sub-exponential decay bound for a large class
of random differential operators (an adaptation to lattices and graphs is straightforward). Unlike
\cite{GK01}, our aim here is to obtain an elementary proof in the simplest situation, without
a more involved bootstrap procedure. In the light of Sections
\ref{sec:SW.ETV}--\ref{sec:MSA.to.DL},
it suffices to work with the resolvents at a fixed energy $E\in I \subset \DR$.

The advantage of the method presented below is that it gives rise to exponential bounds
on the decay of eigenfunctions, while using an exponential sequence of scales,
$L_{k+1} = Y L_k$, as in \cite{Sp88} and in
\cite{GK01}, gives directly only a sub-exponential bound.
(Recall that \cite{GK01} uses several multi-scale analyses to obtain final results, including
\emph{exponential} spectral localization. In \cite{Sp88}, it was indicated
that exponential localization requires scales $L_k \sim L_0^{\alpha^k}$.)

The main idea of the method described below is quite natural. The MSA induction
shows clearly that the exponent $\kappa>0$ in the power-law bound of the form
$\pr{\ball_{L_k}(x) \text{ is \EmS }} \le L_k^{-\kappa}$ grows with the number $K_k$ of allowed
singular $L_{k-1}$-balls inside $\ball_{L_k}(x)$. We allow for a growing
number $K_k\sim L_k^c$, $c\in(0,1)$,  and use an elementary probabilistic bound on such an event,
close in spirit to the Poisson limit theorem. In \cite{GK01}, a similar effect is achieved by a refinement of an idea from \cite{Sp88}: replacing
the sequence of scales $L_k \sim L_0^{\alpha^k}$, $\alpha>1$, by a slower growing
sequence $L_k \sim Y^k L_0$, while keeping uniformly bounded the maximal number $K_k$
of allowed singular cubes.

\subsection{Multiple singular balls: a probabilistic estimate}

\begin{lemma}\label{lem:prob.Poisson}
Suppose that for any ball $\ball_{L_j}(x)\subset\ball_{L_{j+1}}(u)$ one has
$$
\pr{ \ball_{L_j}(x) \text{ is {\rm \EmS}} }\le \eul^{-L_j^\delta},
\;\delta >0.
$$
Let $\cN(\om)$ be the maximal cardinality of collections of pairwise disjoint \EmS balls
$\cC = \{\ball_{L_j}(x_i), i=1, \ldots, \cN\}$. Then,
for $\sigma > \delta$ and $L_0$ is large enough,
$$
\pr{ \cN(\om) \ge L_{j}^{\sigma(\alpha-1)}}
\le \frac{1}{2} \eul^{-L_{j+1}^{\delta}}.
$$
\end{lemma}
\proof
Fix a possible
(unordered) configuration of centers $x_i$ of disjoint \EmS balls, $i=1, \ldots, k$.
Let $N = |\ball_{L_{j+1}}(u)|$.
The number of such configurations for a fixed $k$ is bounded by
$N(N-1)\cdots (N-k+1)/k! \le N^k/k!$, since choosing every center in the sequence $x_1, x_2, \ldots$, excludes at least one possible position for the next center (indeed, many more). For a given configuration, the events $\{\ball_{L_j}(x_i) \text{ is \EmS }\}$ are independent, with probabilities
$
\le p:= \eul^{-L_j^\delta},
$
so
$$
\pr{ \cN \ge n} \le \sum_{k=n}^N \frac{ N^k}{ k! } p^k
\le \frac{(Np)^n}{n!} \sum_{k=0}^\infty  \frac{(Np)^k}{(k+n)!}
 \le \frac{(Np)^n}{n!} e^{Np} \le (Np)^n
$$
for $p < N^{-1}$ and $n\ge 3$.
With $N \le  C L_{j}^{\alpha d}$ and
$n := [L_{j}^{\sigma(\alpha-1)}]$, one has $Np \le \eul^{-L_j^\delta + C \ln L_j}$, thus
$$
\bal
\pr{ \cN \ge n } &\le  (Np)^n
%
 \le
\exp\left\{ -  \left(L_{j}^{\delta} - C \ln L_j \right)
\left( L_j^{\sigma(\alpha-1)  } -1\right)
\right\}
\\
&
\le
 \exp\left\{ - \frac{1}{2} L_{j+1}^\frac{\delta + \sigma(\alpha-1)}{\alpha}
\right\}.
\eal
$$
The condition $\frac{\delta + \sigma(\alpha-1)}{\alpha} > \delta$
is equivalent to the assumed inequality $\sigma>\delta$. Therefore, for some
$\delta'>\delta$ and $L_0$ large enough
$$
\pr{ \cN \ge L_j^{\sigma(\alpha-1)} } \le \eul^{ - \half L_{j+1}^{\delta'} }
\le \frac{1}{2} \eul^{-L_{j+1}^{\delta}}.
\qedhere
$$

\subsection{Decay of $(\ell,q)$-subharmonic functions with "singular" points}

The radial descent bound given by Lemma \ref{lem:subh.2} will require an adaptation.

\begin{definition}\label{def:SH.annuli}
Let $\cG$ be a finite connected graph and $L\ge \ell\ge 0$ two integers and $q\in(0,1)$.
Consider a ball $\ball_L(u)\subsetneq \cG$ and a function $f:\cG\to\DR_+$.
\begin{enumerate}
  \item We say that a point $x\in\ball_L(u)$ is $(\ell,q)$-regular for the function $f$ iff
$$
f(x) \le q \max_{y\in\ball_{\ell + 1}(x)} f(y)
$$
and denote by $\cR(f) (\subset \ball_L(u))$ the set of all regular points of function $f$.
  \item  Given a point
$x\in\ball_L(u)$, let $R(x)$ be the smallest integer
such that $\rS_{R(x)}\subset\cR(f)$; if no such integer exists, we set formally
$R(x) = +\infty$.
  \item  We say that $f$ is
$(\ell,q,\cR)$-subharmonic in $\ball_L(u)$, with $\cR = \cR(f)$, if for any point $x$ with $R(x)<\infty$
and for all $r\ge 0$ such that
$$
\rS_{r+\ell + 1}(u) \subset \ball_{L+1}(u), \quad
\rS_r(u) \subset \cR,
$$
one has
\be\label{eq:SH.annuli.ball}
f(x) \le q \max_{y\in\ball_{r+\ell + 1}(x)} f(y).
\ee
\end{enumerate}
\end{definition}

\begin{lemma}\label{lem:RDB.annuli}
Let a function $f: \cG\to\DR_+$, defined in a finite connected graph $\cG$,
be $(\ell,q,\cR)$-subharmonic in $\ball_L(u)\subsetneq \cG$.
Suppose that
the set $\cR^c = \ball_L(u)\setminus\cR$ is covered by a family
of annuli
$$
\DA = \{A_i, 1\le i \le n \}, \;A_i = \ball_{b_i}(u)\setminus\ball_{a_i}(u),
\;b_i-a_i\le c_i\ell, c_i\in\DN^*,
$$
of total width
$$
w(\DA) = \sum_i(b_i-a_i)\le \sum_i c_i \ell =  C\ell,
$$
with $C\in\DN$ and $2C(\ell+1) < L$.Then
\be
f(u) \le q^{\left\lfloor \frac{L+1 }{\ell + 1}\right\rfloor - 2C} \cM(f,\cG).
\ee
\end{lemma}

\begin{center}
\begin{figure}
\begin{tikzpicture}
\draw[white] (-1,-2) rectangle (12,2);

\draw[thick] (0,0) -- (11,0);
\foreach \i in {0.5, 1.0, ..., 10}
{
  \draw[thin] (\i, -0.1) -- (\i, 0.1);
}
\draw[thick, gray] (1.8, 0.15) -- ++(1.0, 0);
\foreach \i in {1.5, 2.0, 2.5}
{
  \draw[thin, lightgray] (\i, -0.1) -- (\i, 0.1);
}

\draw[line width =3pt, lightgray] (1.5, 0.0) -- ++(1.45, 0);
\node at (2.3, 0.9) (I1) {$J_1=[a_1, b_1]$};
\node at (1.9, 1.3) (c1) {$c_1=2$};
\draw[->] (1.7, 0.6) to [bend left=45] (2.3, 0.25);
\node at (2.2, -0.3) (brace) {$ \underbrace{\,\qquad\qquad\,}$};
\node at (2.3, -0.8) (J1) {$ (2+1)(\ell+1)$};
\node at (2.3, -1.2) (J1) {$ < 2\cdot 2(\ell+1)$};

\draw[thick, gray] (7.4, 0.15) -- ++(1.0, 0);
\draw[line width =3pt, lightgray] (7.0, 0.0) -- ++(1.41, 0);
\foreach \i in {7.0, 7.5, 8.0}
{
  \draw[thin, lightgray] (\i, -0.1) -- (\i, 0.1);
}
\fill (7.2, 0) circle (0.07);
\node at (7.2, 0.1) (ps) {};
\node at (7.3, -0.5) (s) {$s$};
\node at (8.3, 1) (hr) {$h(s)\le q\, h(r')$};
\fill (8.5, 0) circle (0.07);
\node at (8.5, 0.1) (prl) {};
\node at (8.9, -0.5) (rl) {$r'\in\cR$};
\draw[->] (ps.north) to [bend left=90] (prl.north);

\fill (5.0, 0) circle (0.07);
\node at (5.0, 0.1) (pra) {};
\node at (5.0, -0.5) (ra) {$r$};
\node at (5.450, -1.2) (hr1) {$\overbrace{r'=r+(\ell+1)}$};
\draw[->] (5.5, -0.8) to [bend right=30] (5.5, -0.2);
\fill (5.5, 0) circle (0.07);
\node at (5.5, 0.1) (pra1) {};
\node at (5.5, 0.8) (hra1) {$h(r) \le q\, h(r')$};
\draw[->] (pra.north) to [bend left=90] (pra1.north);

\node at (8.5,-1.5) (J2) {$J_2$};
\draw[->] (8.3,-1.3) to [bend left=30] (7.7,-0.2);

\end{tikzpicture}
\caption{Example for the proof of Lemma \ref{lem:RDB.annuli}.
Recursion for $r\in\cR$ (step of length $\ell+1$) and for $s\in\cR^{\rc}$
(step of length $\le |J_2|$).}
\end{figure}
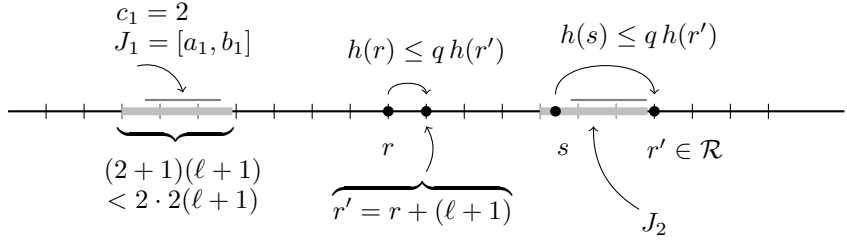
\end{center}

\proof
Divide the integer interval $[0,L+1]\cap \DN$ into
$N:=\left\lfloor \frac{L+1 }{\ell + 1}\right\rfloor$ intervals of the form
$I_j =[j(\ell+1), (j+1)(\ell+1)-1]$, $j = 0, 1, ..., n-1$,  and, eventually, a remainder which will be unused in the argument. Call an interval $I_j$ good if $\rS_{j(\ell+1)}\subset\cR$. Since $\ball_L(u)\subsetneq \cG$, the sphere $\rS_{L+1}(u)$ is non-empty. The radial projection
$$
\ball_{L+1}(u)\ni x\mapsto \rd(u,x)\in[0,L+1]
$$
maps an annulus of width $c_i\ell$ onto an interval of length $c_i\ell$, covered by at most
$c_i+1\le 2c_i$
adjacent intervals of the form $I_j$. Therefore, the entire set $\cR^c$ is radially mapped onto a
subset of $[0,L]$ covered by a family of at most $2C$ intervals $I_j$. Respectively, at least
$K=N - 2C$ intervals $I_{j_1}, I_{j_2}, \ldots, I_{j_K}$ must be good,
and it follows from the hypotheses that $K\ge 1$.
Further, let $h(r) := \max_{x\in\ball_r(u)} f(x)\ge 0$, $r\in[0,L+1]$.
Note that if
$x\in\ball_{j(\ell+1)}(u)$ and $z\in\ball_{\ell+1}(x)$, then
$z\in\ball_{(j+1)\ell+1}(u)$. For $x\in\cR(f)$ with $\rd(u,x)=:r\in I_{j_i}$ one can apply
\eqref{eq:lqsubh}, so for any $1 \le i \le K$, one has
\be\label{eq:RDS.1}
\bal
h( j_i(\ell+1)) & = \max_{x\in\ball_{ j_i(\ell+1) }(u)} f(x)
\le q \max_{z\in\ball_{ (j_i+1)(\ell+1) }(u)} f(z)
\\
& = q h((j_i+1)(\ell+1)).
\eal
\ee
The value $i=K$, i.e., the index $j_K$, is admissible here, since
$$
(j_{K}+1)(\ell+1) \le K(\ell+1) + 2C(\ell+1) \le L+1,
$$
and the set $\rS_{L+1}(u)$ is nonempty, by assumption.
For $i=K$, we have, by monotonicity of the function $h$, the following upper bound on the RHS
of \eqref{eq:RDS.1}:
$$
q h((j_i+1)(\ell+1)) \le q h(L+1) \le q \| f \|_\infty,
$$
so by backward recursion $i=K-1, ..., 1$, one obtains in $K-1$ steps:
$$
f(u) = h(0) \le h(j_1(\ell+1)) \le q^K\, \|f\|_\infty
\le q^{\left\lfloor \frac{L+1 }{\ell + 1}\right\rfloor - 2C}\, \|f\|_\infty.
\qedhere
$$

Definition \ref{def:SH.annuli} is tailored so as to suit the following analog of
Lemma \ref{lem:GF.is.SH.1}.

\begin{lemma}\label{lem:GF.EF.SH}
Consider a finite graph $\cG$, a DSO $H_\cG$ and a ball $\ball_{L_{k+1}}(u)\subsetneq\cG$.
Fix some $E\in\DR$, assume that $\ball_{L_{k+1}}(u)$ is \CNR, and let $\cR = \cR(\cG,E)$ be the set of points $x\in\cG$ such that the ball $\ball_{L_k}(x)\subset\cG$ is $(E,m)$-NS. Then
for any $y\in\cG\setminus\ball_{L_{k+1}}(u)$, the function
  $x\mapsto \big|G_{\ball_{L_{k+1}}(u)}(x,y;E)\big|$ is $(L_k,q,\cR)$-subharmonic in $\ball_{L_{k+1}}(u)$, with
      $q = e^{ m(1 + \half L_k^{-\tau})L_k}$, if $L_0$ is large enough.

\end{lemma}

\proof
If $x\in\cR$, then, by the \EmNS property of $\ball_{L_{k}}(x)$,
$$
f(x) \le  \eul^{-\gamma(m, L_k)L_k} \cM(f, \ball_{L_{k}+1}(x)).
$$
For $x\in\cR^{\rc}$ with $R(x)<\infty$ (the radius $R(x)$ is defined
as in Definition \ref{def:SH.annuli}), we have, by \ENR property of the ball
$\ball_{R(x)-1}(u)$, stemming from the assumed \CNR property of
$\ball_{L_{k+1}}(u)$,
\begin{align*}
f(x) &\le |\pt \ball_{L_{k}}(x)| \;\| G_{\ball_{R(x)-1}(u)}(x,v;E)\|
\, \max_{\rd(u, v') = R(x)} | G_{\ball_{L_{k+1}}(u)}(v',y;E)|
\\
& \le \eul^{L_k^\beta} \; \max_{\rd(u, v') = R(x)} | G_{\ball_{L_{k+1}}(u)}(v',y;E)|
\\
\intertext{and since $S_{R(x)}(u)\subset\cR$,}
&< \eul^{L_k^\beta} \max_{v'\in S_{R(x)}(u)}
\eul^{-\gamma(m, L_k)L_k} \cM(f, \ball_{1+L_{k}}(v'))
\\
& \le \eul^{-\gamma(m, L_k)L_k + L_k^\beta} \cM(f, \ball_{R(x) +L_{k}+1}(u)) .
\end{align*}
For $L_0$ large enough and $\tau<1-\beta$, one has
$\gamma(m, L_k)L_k - L_k^\beta \ge m(1 + \half L_k^{-\tau})L_k$.
\qedhere

\subsection{Scaling with sub-exponential bounds}

From this point on, we fix the key parameters used in the scale induction as follows:

\begin{center}
\begin{tabular}{|l|l|}
  \hline
  $\alpha =\tabhigh{\frac{4}{3}}$ & $\beta = \frac{1}{3}$ \\
  \hline
   $\delta = \frac{1}{4} < \beta$ &  $\sigma = \tabhigh{\frac{1}{3} > \delta}$\\
  \hline
  $\rho = \tabhigh{ \frac{1}{3}}$ & $\tau=\frac{1}{8}=\frac{1}{2}\big(\rho-\sigma(\alpha-1)\big)$
  \\
  \hline
\end{tabular}
\end{center}
\par
\vskip3mm

\begin{definition}
A ball $\ball_{L_{k+1}}(x)$ is called $E$-tunneling (\ET) if, for some $E\in I$, it contains a collection of more than $L_k^{\sigma(\alpha-1)}$ pairwise disjoint \EmS balls of radius $L_k$, and $E$-non-tunneling (\ENT), otherwise.
\end{definition}

\begin{lemma}\label{lem:NT.CNR.implies.NS.bootstrap}
If a ball $\ball_{L_{k+1}}(u)$ is \ENT and \CNR,
then it is \EmNS.
\end{lemma}

\proof
Fix two points $x,y\in\ball_{L_{k+1}}(u)$ with
$\rd(x,y)\ge  L_{k}^{1+\varrho}$ and let
$R = \rd(x,y) - L_k$.
By Lemma \ref{lem:GF.EF.SH}, the function $f:z\mapsto G_{\ball_{L_{k+1}}}(x,z)$  is $(L_k, q,\cS)$-subharmonic in $\ball = \ball_{R}(x)$, with $q=\eul^{-m(1+\half L_k^{-\tau})L_k}$,
and the \ENT assumption guarantees that $\cS$ can be covered by at most
$L_k^{\sigma(\alpha-1)}$ balls of radius $2L_k$, hence by a collection $\DA$ of annuli
centered at $u$ of total width $w(\DA)\le L_k^{\sigma(\alpha-1)} \cdot 4L_k$.
Therefore,
\be\label{eq:psi.psi.q}
f(x)
\le q^{\frac{R- w(\DA) - 2L_k}{L_k + 1}}  \|f\|_\infty
\le q^{\frac{R- 5 L_k^{1+\sigma(\alpha-1)} }{L_k + 1}} \|f\|_\infty.
\ee
Recall that
$\varrho - \sigma(\alpha - 1)=2\tau$, so
$$
R - 5 L_k^{1+\sigma(\alpha-1)}
> R\left( 1 - 5 L_k^{\sigma(\alpha-1)  -\varrho } \right)
 = R\big( 1 - 5 L_k^{- 2\tau} \big).
$$
Put this lower bound into \eqref{eq:psi.psi.q} and write, with the convention $-\ln 0 = +\infty$:
\begin{align*}\label{eq:proof.NS}
-\ln f(x)
&\ge
-\ln
\Bigg\{ \left(
e^{ -m(1+ \half L_{k}^{-\tau})L_{k} }
\right)^{\frac{R\left( 1 - 5 L_k^{- 2\tau} \right)}{L_{k}+1}}
e^{ L_{k+1}^\beta} \Bigg\}
\\
& \ge mR\left( \left(1+ {\textstyle \half } L_{k}^{-\tau} \right)
    {\textstyle \frac{L_{k}}{ L_{k} + 1} } \left( 1 - 5 L_k^{- 2\tau} \right)
    - \frac{  L_{k+1}^{\beta}}{mR} \,\right)
\intertext{ (with $m\ge 1$,
$R = \rd(x,y) - L_k > L_k^{1+\varrho}-L_k > \half L_k^{1+\varrho}$\, and $L_0$ large enough)}
 &\ge
 mR \left( (1 + {\textstyle \half }L_{k}^{-\tau})  \left(1 - 6 L_k^{- 2\tau} \right)
    - 2 L_{k}^{-1-\varrho + \alpha\beta} \right)
\\
\intertext{ (use
$\rho = \frac{1}{3} $, $\alpha\beta = \frac{4}{9}$
$\Rightarrow$ $1 + \rho - \alpha\beta = \frac{8}{9} > 2\tau $) }
& \ge  mR \left(1 +{\textstyle \half } L_{k}^{-\tau} \right)  \left(1 - 7 L_k^{- 2\tau} \right)
\ge m\left(1+ {\textstyle \quart} L_{k}^{-\tau} \right)\, \rd(x,y)
\\
& \ge \gamma(m,L_{k+1}) \,\rd(x,y).
\qedhere
\end{align*}

\vskip4mm

Consider the following property which we will prove by induction for all $k\in\DN$:
\medskip

\SS{k, E}:
For  any ball $\ball_{L_{k}}(x)\subset\cZ$, one has
\be\label{eq:thm.ind}
\pr{  \ball_{L_k}(x) \text{ is \EmS} }
\le \eul^{-L_k^\delta} .
\ee

\begin{theorem}\label{thm:subexp.ind.loc}
{\rm \SS{k,E}} implies {\rm \SS{k+1,E}}.
\end{theorem}

\proof
Denote by $N(\ball_{L_{k+1}}(x))$ the maximal cardinality of collections of pairwise disjoint
\EmS balls of radius $L_{k}$ inside $\ball_{L_{k+1}}(x)$.
Introduce the events
$$
\bal
\cB_{k+1} &= \left\{ N(\ball_{L_{k+1}}(x)) \ge L_{k}^{\sigma(\alpha-1)} \; \right\},
\\
\cE_{k+1} &= \left\{  \text{ $\ball_{L_{k+1}}(x)$ is \PR } \right\}.
\\
\eal
$$
By Lemma \ref{lem:NT.CNR.implies.NS.bootstrap},
$\{ \om:\, \text{ $\ball_{L_{k+1}}(x)$  is \EmS } \} \subset \cE_{k+1} \cup \cB_{k+1}$. By
Wegner estimate \eqref{eq:Wegner}, using $\beta >\delta$, we have
$$
\textstyle
\pr{  \cE_{k+1} } \le  \frac{1}{2} \eul^{ - L_{k}^\beta}
\le  \frac{1}{2} \eul^{ - L_{k}^\delta}
$$
so the
claim follows from Lemma \ref{lem:prob.Poisson} saying that
$\pr{\cB_{k+1}} \le \frac{1}{2} \eul^{ - L_{k}^\delta}$.
\qedhere

\vskip 2mm

For the initial scale bound \SS{0} (indeed, \emph{any} desired probabilistic bound,
cf. \cite{DK89})
to hold true, it suffices to pick $|g|$ large enough.
Therefore, we come by induction to the following

\begin{theorem}\label{thm:fixed.subexp}
Assume that the random potential fulfills the regularity
condition\footnote{As was said, the Lipshitz continuity condition \eqref{eq:Lipshitz}
can be relaxed to the H\"{o}lder continuity.}
\eqref{eq:Lipshitz}. Then for all $|g|$ large enough, the property \SS{k} holds true
for all $k\ge 0$.
\end{theorem}

\begin{remark}
A more tedious (but elementary) parametric analysis shows that one can actually
choose the exponent $\delta\in(0,1)$ arbitrarily close to $1$, thus getting a
sub-exponential decay very close to the exponential one. Indeed,
the complete set of requirements for the scaling parameters
is as follows:
$$
\bal
0 &<  \rho < \alpha - 1
\\
0 &<  \sigma < \frac{\rho}{ \alpha - 1}, \quad
%
0 < \delta < \min\{\beta, \sigma  \}
\\
0 &< 2\tau  = \min\{ \rho - \sigma(\alpha - 1), 1 + \rho - \alpha\beta \}
\eal
$$
A direct inspection shows that for any $\eps\in(0, \half)$, one can set, e.g.,
\begin{center}
\begin{tabular}{|l|l|}
  \hline
  $\alpha =\tabhigh{ 1 + 4\eps }$ & $\beta = 1- \eps$ \\
  \hline
   $\delta = 1 - 2\eps  < \beta$ &  $\sigma = \tabhigh{1  - \eps > \delta}$\\
  \hline
  $\rho = \tabhigh{4\eps - 2\eps^2 }$ & $\tau=\eps^2=\frac{1}{2}\big(\rho-\sigma(\alpha-1)\big)$
  \\
  \hline
\end{tabular}
\end{center}
\end{remark}
\par
\vskip3mm

\section{On a direct approach to the decay of eigenfunctions }
\label{sec:DSA}

We have seen that the fixed-energy analysis of resolvents implies the spectral and dynamical
localization. On the  other hand, it is also possible to adapt the approach from the
Spencer's work \cite{Sp88} to a direct analysis of eigenfunctions in arbitrarily large finite
balls. For single-particle models, such an adaptation has been proposed
in our recent paper \cite{C12}. The new scaling scheme can be summarized as follows:

\begin{itemize}
  \item The key notion becomes that of an $m$-localized ball. We say that a ball
  $\ball_L(u)$ is $m$-localized if the eigenbasis $\{\psi_i\}$
  of the operator $H_{\ball_L(u)}$ fulfills the following condition: for all points $x,y\in\ball_L(u)$
  with $\rd(x,y) \ge L^{7/8}$,
$$
\sum_{\lam_i\in\Sigma(H_{\ball_L(u)})} \; |\psi_i(x)\, \psi_i(y)|
\le \eul^{-\gamma(m,L)\rd(x,y)}.
$$
%
%
  \item A ball $\ball_{L_{k+1}}(u)$ is called $E$-completely non-resonant (\CNR) if it is \ENR
  and contains no \ER ball of radius $\ge L_k$.
  \item It is readily seen from the eigenfunction expansion of the resolvent that an
  $m$-localized ball which is $E$-NR must be
  \EmNS.

  \item A direct analog of Lemma \ref{lem:NR.NT.implies.NS} is still valid.

  \item Consider the bounds (which we will denote by \LOC{k}, $k\ge 1$) of the form
\be\label{eq:mloc.0}
  \pr{ \ball_{L_k}(u) \text{ is $m$-localized }} \ge 1 - L_k^{-\kappa(1+\theta)^k}.
\ee
  The initial scale bound \LOC{1}, for any $m\ge 1$, $L_0\ge 1$ and $|g|$ large enough,
  is easily obtained by elementary perturbation theory for self-adjoint operators (which are finite-dimensional here) with simple  spectrum.
  The induction step (\LOC{k} $\Rightarrow$ \LOC{k+1}) is performed as follows.
      \begin{itemize}
        \item Assume \LOC{k} and consider a ball $\ball_{L_{k+1}}(u)$. If it is not
        $m$-localized, then by an analog of Lemma \ref{lem:NR.NT.implies.NS}, for some $E\in\DR$
        it must contain two disjoint \EmS balls $\ball_{L_k}(x)$, $\ball_{L_k}(y)$. One can easily infer from the Wegner estimate that with probability
        $\ge 1 - \eul^{-L_{k+1}^{\beta/2}}$, for any $E\in\DR$
        either $\ball_{L_k}(x)$ or $\ball_{L_k}(y)$ is \CNR.
        \item
        If $\ball_{L_k}(x)$ is \EmS and \CNR, it must contain a pair of disjoint
        $m$-non-localized balls of radius $L_{k-1}$. By virtue of the inductive assumption
        \LOC{k}, the probability of the latter event is bounded by
        $C L_{k-1}^{2d\alpha^2 - 2\kappa(1+\theta)^{k-1}}$.
        \item For $\alpha\in(1, \sqrt{2})$, $\kappa > \frac{2\alpha^2 d}{2 - \alpha^2}$
        and an appropriately chosen $\theta>0$, the above bounds imply \LOC{k+1}.
      \end{itemize}
\end{itemize}

The property \LOC{k}, proven for all $k\ge 0$, is already a form of localization of
eigenfunctions. In addition, it implies the usual variable-energy MSA bounds, hence,
the strong dynamical localization and an exponential decay of eigenfunctions.

It is to be emphasized that the main analytic tool of the simplified MSA remains the elementary
Lemma \ref{lem:subh.1} (and Lemma \ref{lem:subh.2} easily stemming from it). As was mentioned earlier, the idea of the "two-sided" estimates of the Green functions (and, similarly,  eigenfunctions)
goes back to Spencer's work \cite{Sp88}.

\section*{Conclusion}

We have shown that the fixed-energy probabilistic analysis of random Anderson-type
Hamiltonians implies, in a fairly general and elementary way, stronger manifestations of
the Anderson localization phenomenon, viz.: spectral and strong dynamical localization.
The new method, going in the same direction as the well-known Simon--Wolf criterion
of localization, applies also to multi-particle systems (as will be shown in our forthcoming work)
for which no analog of the Simon--Wolf approach has been developed so far. Moreover,
combined with
the simplified Germinet--Klein argument, it gives rise to the strong dynamical localization, not
only spectral localization. Therefore, a very simple scaling procedure going back to Spencer's
work
\cite{Sp88}, as well as its counterpart for interacting multi-particle systems, results in a simple proof
of strong dynamical localization for a large class of models.

\section*{Acknowledgements}
It is a pleasure to  thank Tom Spencer and the Institute for Advanced Study, Princeton,
for their hospitality
during my visit to the IAS in March, 2012, and for numerous stimulating discussions;  Ya. G. Sinai
for fruitful discissions of the spectral properties of almost periodic operators;
Sasha Sodin for fruitful discussions of the works \cite{ETV10}--\cite{ESS12}; G\"{u}nter Stolz,
Yulia Karpeshina, Roman Shterenberg and the University of Alabama at Birmingham for their
hospitality during my visit to the UAB in March, 2012, and for numerous  discussions
of localization  phenomena in multi-particle systems and localization/delocalization in almost periodic media.


\end{document}